\documentclass[twocolumn]{aastex631}

\newcommand{\chandra}{\textit{Chandra}}

\newcommand{\xmm}{\textit{XMM-Newton}}

\newcommand{\nustar}{\textit{NuSTAR}}

\newcommand\zwcl{ZWCL 1856.8}

\usepackage{multirow}
\usepackage{wrapfig} 
\shorttitle{\zwcl}
\shortauthors{T\"{u}mer et al.}

\begin{document}

\title{\zwcl~: A rare double radio relic system captured within \nustar~and \chandra~field of view}

\correspondingauthor{Ay\c{s}eg\"{u}l T\"{u}mer}
\email{aysegultumer@gmail.com}

\author[0000-0002-3132-8776]{Ay\c{s}eg\"{u}l T\"{u}mer}
\affiliation{Kavli Institute for Astrophysics and Space Research, Massachusetts Institute of Technology, 77 Massachusetts Avenue, Cambridge, MA 02139, USA}
\affiliation{Department of Physics \& Astronomy, The University of Utah, 115 South 1400 East, Salt Lake City, UT 84112, USA}

\author[0000-0001-9110-2245]{Daniel R. Wik}
\affiliation{Department of Physics \& Astronomy, The University of Utah, 115 South 1400 East, Salt Lake City, UT 84112, USA}

\author[0000-0002-4962-0740]{Gerrit Schellenberger}
\affiliation{Center for Astrophysics $|$ Harvard \& Smithsonian, 60 Garden St., Cambridge, MA 02138, USA}

\author[0000-0002-3031-2326]{Eric D. Miller}
\affiliation{Kavli Institute for Astrophysics and Space Research, Massachusetts Institute of Technology, 77 Massachusetts Avenue, Cambridge, MA 02139, USA}

\author{Marshall W. Bautz}
\affiliation{Kavli Institute for Astrophysics and Space Research, Massachusetts Institute of Technology, 77 Massachusetts Avenue, Cambridge, MA 02139, USA}

\begin{abstract}

Observations of galaxy cluster mergers provide insights on the particle acceleration and heating mechanisms taking place within the intracluster medium. Mergers form shocks that propagate through the plasma, which result in shock/cold fronts in the X-ray, and radio halos and/or relics in the radio regime. The connection between these tracers and the mechanisms driving non-thermal processes, such as inverse Compton, are not well understood. \zwcl~is one of the few known double radio relic systems that originate from nearly head-on collisions observed close to the plane of the sky. For the first time, we study \nustar~and \chandra~observations of such a system that contains both relics within their field of view. The spectro-imaging analyses results of the system suggest weak shock fronts with $\mathcal{M}$ numbers within 2$\sigma$ of the radio derived values, and provide evidence of inverse Compton emission at both relic sites. Our findings have great uncertainties due to the shallow exposure times available. Deeper \nustar~and \chandra~data are crucial for studying the connection of the radio and X-ray emission features and for constraining the thermal vs.\ non-thermal emission contributions in this system. We also present methods and approaches on how to investigate X-ray properties of double relic systems by taking full advantage of the complementary properties of \nustar~and \chandra~missions.

\end{abstract}

%% https://astrothesaurus.org

\keywords{X-rays: galaxies: clusters --- galaxies: clusters: individual (\zwcl), intracluster medium --- radiation mechanisms: non-thermal, thermal --- shock waves}

\section{Introduction} \label{sec:intro}

According to large scale structure formation scenarios, clusters of galaxies are hierarchically  formed by the merger of smaller scale structures. Galaxy cluster mergers are the most energetic ($10^{64-65}$~ergs) processes in the universe, which drive shocks and turbulence into the intracluster medium (ICM). As dark matter halos and galaxy velocity distributions revirialize relatively slowly, the kinetic energy of the merger is transformed more quickly within the collisional hot gas, driving shock and cold fronts that heat and mix the thermal gas while also (re)accelerating electrons and cosmic rays through Fermi-like acceleration processes \citep[see, e.g.][]{brunetti14}. 

Shock fronts in the X-ray regime are evidenced by sharp surface brightness (SB) discontinuities accompanied by a temperature drop from the upstream to the downstream of the front. Merger shocks are characterized by a Mach number of~$<$~3 \citep[see, e.g.][]{gabici03,ryu03}. Cold fronts, on the other hand, are due to the motion of cooler subcluster in the ambient gas with higher entropy with respect to the subcluster. Cold fronts exhibit similar SB discontinuity behaviour like the shock fronts, but the temperature drop direction is reversed; the temperature ahead of the cold front is higher than the temperature behind \citep{markevitch07}.

In addition to shock and cold fronts, mergers of galaxy clusters form extended radio structures in the form of radio halos and relics. Galaxy clusters hosting these structures provide information on the astrophysical particle acceleration mechanisms as well as non-thermal processes that take place within the ICM \citep{finner21}. Radio halos are thought to be formed by the turbulence introduced by merger shocks, and are found in the central regions of these systems. Radio relics (radio shocks), on the other hand, are anticipated to trace large-scale shock waves propagating through the ICM \citep{degasperin14}. Relics usually reside in galaxy cluster periphery, since half of the virial radius from the cluster center is where the kinetic energy dissipated in merger shocks peaks \citep{vazza12}. Further details on the particle acceleration mechanisms forming extended radio structures and their efficiencies are provided by \citet{bonafede12, weeren19}.

Double radio relic formation, which are nearly equidistant from the cluster center, are quite rare. These structures are thought to have formed as a result of a single merger event and merely 12 double relics are identified in literature \citep{weeren19}. This scarcity is due to the merger geometry; i.e., the merger has to take place near the plane of the sky for both relics to be visible to the observer, in addition to the requirement of nearly head-on collisions of similar mass systems. Since each of the double relic pair trace the same single merger event, they provide the advantage of putting tighter constraints on the merger scenarios \citep{finner21} and particle acceleration mechanisms. 

The coincidence of X-ray detected shock fronts and radio detected relics in merger systems is evidenced by multiple studies, but require further confirmations \citep{weeren19}. To confirm shock fronts at relic locations, where the S/N is low due to the large distances from the X-ray peaks, deep observations with high spatial resolution are required to locate the surface brightness discontinuities, as well as wide X-ray bandwidth to constrain the temperature drops across these edges. These confirmations are extremely important to shed light on the particle acceleration and ICM heating mechanisms.

Wide X-ray band also provide key insights into the conditions and mechanisms that are operating in the ICM with the capability of disentangle the thermal and non-thermal emission contributions. The same relativistic electrons producing the synchrotron-powered radio halos also upscatter cosmic microwave background (CMB) photons to X-ray energies via inverse Compton (IC) process. The ratio of fluxes between the synchrotron and this IC emission is simply the ratio of the energy densities of their respective radiation fields, i.e.; the magnetic field strength $B$ and the CMB, respectively. Since the latter is well known, an IC detection or upper limit leads to an estimate or lower limit on the volume-averaged value of $B$, a quantity that is poorly constrained in galaxy clusters in general, which has significant implications on the assumptions used for galaxy cluster mass estimates \citep[see, e.g.][]{Biffi16,Sarazin16}. IC emission in galaxy clusters requires better constraints to confirm that the dynamical role of magnetic fields in clusters are negligible, as is currently assumed in simulations and mass scaling relations used in cosmology (e.g., \citep{vikhlinin09}). 

To constrain the IC emission within the ICM, hard X-ray data are required that enable disentangling the thermal bremsstrahlung and relatively weak non-thermal emission features in the ICM. Narrow X-ray band is incapable of disentangling the tail of IC emission that extends beyond the thermal bremsstrahlung turn over at higher energies \citep{wik14}.

\cite{degasperin14} discovered a double radio relic system in ZWCL 1856.8+6616 (\zwcl~herein) at z = 0.304. The cluster is also known as PSZ1(PSZ2) G096.89+24.17, which is detected through the Sunyaev-Zeldovich (SZ) effect by Planck Survey \citep{planck11} whose mass is estimated to be \textit{M$_{500}$}~$\simeq$~4.7$\times$10$^{14}$ \textit{M$_{\odot}$} \citep{planck16}. Using ROSAT X-ray flux, the X-ray luminosity is estimated to be \textit{L$_{x}$}~=~3.7~$\times$10$^{44}$ erg/s \citep{degasperin14}. \citet{degasperin14} estimate the average cluster temperature to be \textit{kT}~$\sim$~4~keV using \textit{L$_{x}$}~-~\textit{T$_{x}$} scaling relations provided by \citet{pratt09}. The cluster is also observed with \xmm~($\sim$12 ks, Observation ID~:~0723160401), \chandra~($\sim$42 ks, Observation ID~:~19752) and most recently with \nustar~($\sim$30 ks, Observation ID~:~70801003002). 

\chandra~and \xmm~data show a double peaked cluster structure driven by the merging subclusters \citep{finner21,jones21}. Using Monte Carlo Merging Analysis Code (MCMAC; \citep{dawson13}) \citet{finner21} suggest that \zwcl~ is at an early stage merger; returning from the first apocenter. A weak-lensing study by \citet{finner21} report \textit{M$_{200}$}=2.4$\times$10$^{14}$ \textit{M$_{\odot}$} for the system, and defines the system as a major merger with a near 1:1 mass ratio. They assign masses of \textit{M$_{200}$}=1.2$\times$10$^{14}$ \textit{M$_{\odot}$} and \textit{M$_{200}$}=1.0$\times$10$^{14}$ \textit{M$_{\odot}$} to the north and south subclusters, respectively. While they report a global temperature of \textit{kT}=3.7~keV and \textit{kT}=4.3~keV for the northern subcluster using \xmm~data, they cannot constrain the temperature for the southern subcluster.

Full-polarization WSRT radio observation study at 1.4 GHz by \citet{degasperin14} report the linear sizes of the relics to be $\sim$0.9 and $\sim$1.4 Mpc for the northern and the southern relics, respectively in addition to the hints of a central radio halo. They cannot report radio spectral indices due to the insufficient data. \citet{jones21} report the sizes as $\sim$0.9 and $\sim$1.5 Mpc using LOFAR data at 140 MHz. They report integrated (injection) radio spectral indices of $\alpha_{int(inj)}$=-0.95 (-0.87) and $\alpha_{int(inj)}$=-1.17 (-0.97) for the northern and southern relics respectively. Assuming the shock fronts lie at the outer edge of the relics, \citet{jones21} find spectral steepening towards the cluster center due to the downstream synchrotron and IC losses. They report that the Mach number derived from the integrated spectral indices, $\mathcal{M}_{int}$, is unconstrained for the northern relic and is $\mathcal{M}_{int}=3.6\pm{0.7}$ for the southern relic. Whereas Mach numbers derived from the injection spectral indices, $\mathcal{M}_{int}$, are $\mathcal{M}_{inj}=2.5\pm{0.2}$ and $\mathcal{M}_{inj}=2.3\pm{0.2}$ for the northern and southern relic, respectively. Guided by the low-resolution LOFAR image, they also add that the extended emission lying to the east of the cluster in the region between the relics, may be a candidate radio halo. In addition, strong polarization in the northern relic and weak polarization in the southern relic points to a slightly tilted view from the plane of the sky \citep{degasperin14}. 

At the cluster redshift, \nustar~and \chandra~were able to observe both of the radio relics within one field of view (FOV). This was the main motivation behind requesting \nustar~follow-up observations for obtaining hard X-ray information on the system. To elaborate; when both the regions of interest and the bright galaxy cluster centers do not fall within the same \nustar~FOV, scattered light becomes difficult to tackle with, since bright sources near but out of the \nustar~FOV causes energy and position dependant emission contamination within the FOV \citep{tumer22}. Hence, this \nustar~observation is unique in a way that provides hard X-ray double relic data without scattered light contamination.

In this paper, we investigate the X-ray emission features of this merger via spectro-imaging analyses of \nustar~and \chandra~data. The paper is organized as follows: observations, data reduction processes and the background assessment of the \nustar~and \chandra~data are presented in Section~\ref{sec:reduction}. In Section~\ref{sec:analysis}, the spectro-imaging analyses are described and their results are presented. We discuss our findings in Section~\ref{sec:discussion}, and conclude our work in Section~\ref{sec:future} including the plan for future work.

Throughout this paper, we assume the $\Lambda$CDM cosmology with {\it H$_{0}$} = 70 km s$^{-1}$ Mpc$^{-1}$, $\Omega_{M}$ = {\tt 0.27}, $\Omega_{\Lambda}$ = {\tt 0.73}. According to these assumptions, at the cluster redshift, a projected intracluster distance of 100 kpc corresponds to an angular separation of $\sim$22~$\arcsec$. All uncertainties are quoted at the 68\% confidence levels unless otherwise stated.

\section{Observations and data reduction} \label{sec:reduction}
\subsection{\nustar}
In this work, we use \nustar~\citep[Nuclear Spectroscopic Telescope Array;][]{harrison13} observations of \zwcl\ with Observation ID~:~70801003002 that took place in 2022 using both focal plane modules, i.e, FPMA and FPMB, for a $\sim$30~ks of exposure time for each module.

\begin{figure}[h!]
\centering
\includegraphics[width=80mm]{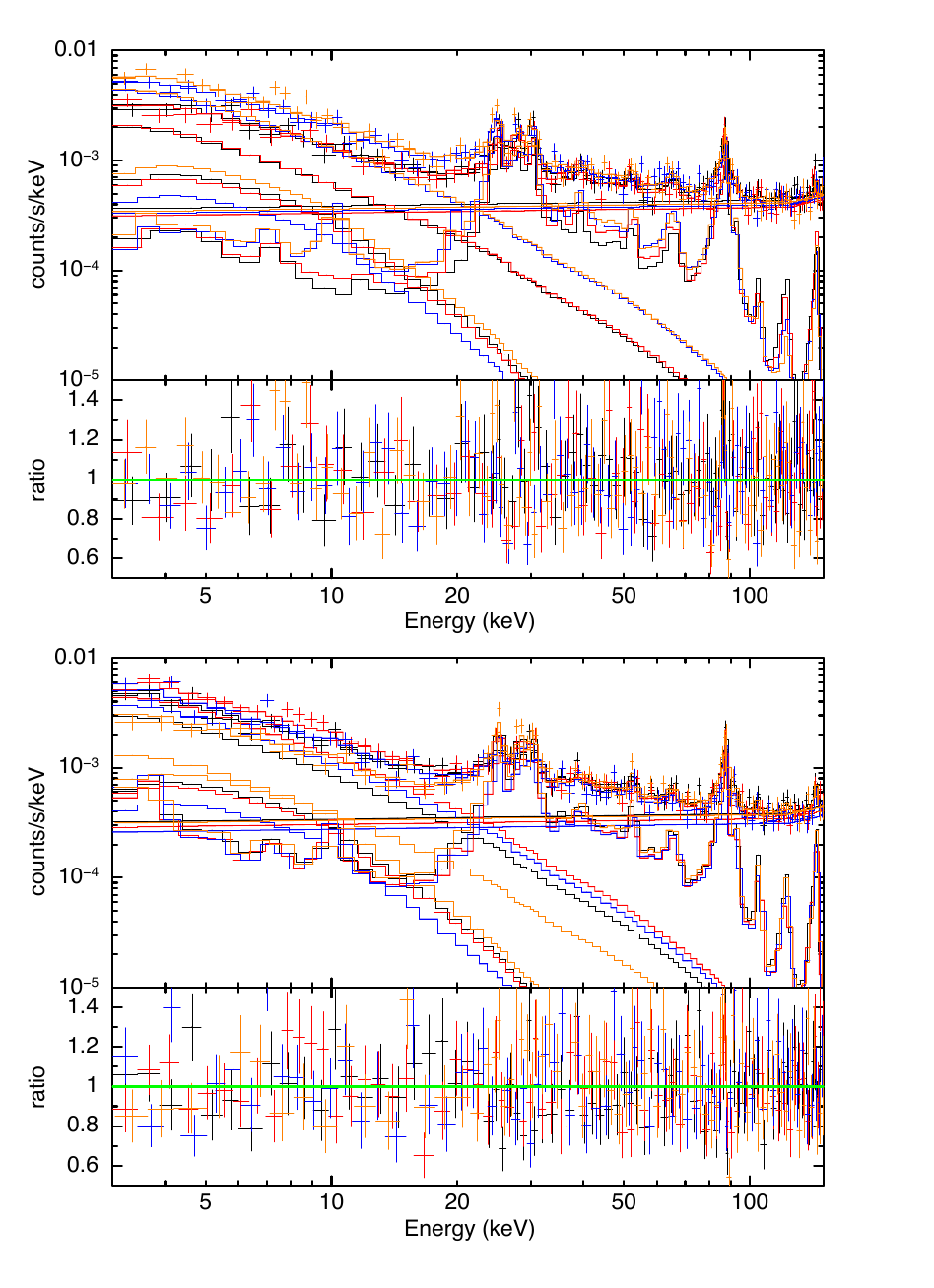}
\caption{Joint-fit of background and cluster emission of \nustar\ FPMA (\textit{upper panel}) and FPMB (\textit{lower panel}). Each color represents a region selected for the background fit. For plotting purposes, adjacent bins are grouped until they have a significant detection at least as large as 8$\sigma$, with maximum 12 bins. \label{fig:nuskybgd}}
\end{figure}

In order to filter the data, standard pipeline processing using HEASoft (v.~6.28) and NuSTARDAS (v.~2.0.0.) tools were used. To clean the event files, the stage 1 and 2 of the NuSTARDAS pipeline processing script {\tt nupipeline} were utilized. Regarding the cleaning of the event files for the passages through the South Atlantic Anomaly (SAA) and a ``tentacle"-like region of higher activity near part of the SAA, instead of using SAAMODE=STRICT and TENTACLE=yes calls, we produced light curves and manually removed increased count rate intervals using 100 s time bins to create good time intervals (GTIs) without fully discarding the passage intervals.

This new set of GTIs were then reprocessed with {\tt nupipeline} stages 1 and 2, and cleaned images were generated at different energy bands with XSELECT. We used {\tt nuexpomap} to create exposure maps in the 3.0~--~8.0 keV and 8.0~--~15.0 keV bands accounting for vignetting. To produce the corresponding spectra for the regions of interest as well as the corresponding Response Matrix Files (RMFs) and Ancillary Response Files (ARFs), stage 3 of {\tt nuproducts} pipeline were used.

\begin{figure*}
\centering
\includegraphics[width=140mm]{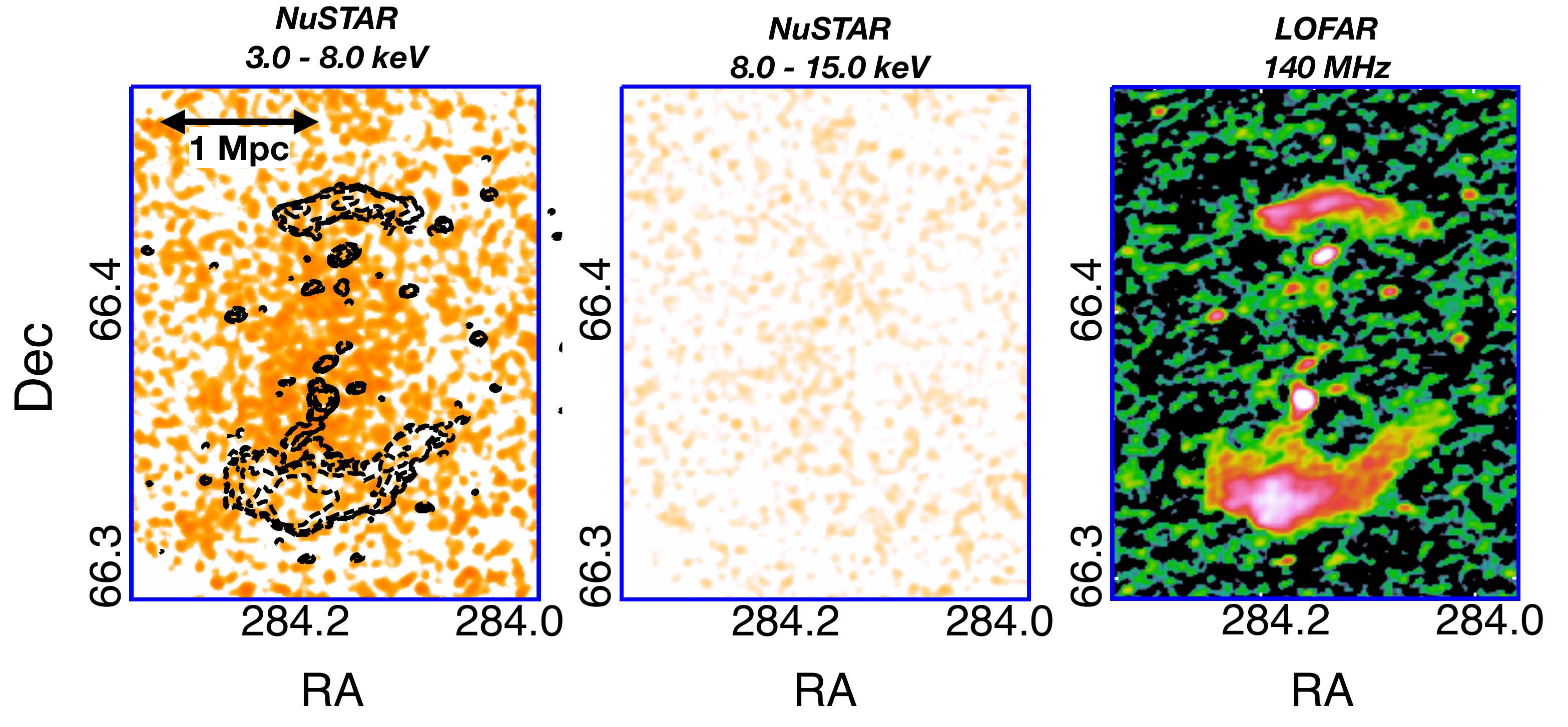}
\caption{Background subtracted, exposure corrected, \nustar~images of \zwcl~smoothed with a Gaussian of r=5 pixels (1 pixel subtending 12$\arcsec$.3) in the soft {\it left} and hard bands {\it middle}, and LOFAR image at 140 MHz (resolution: 13$\arcsec$ $\times$ 7$\arcsec$, RMS noise: 127 $\mu$Jy/beam) {\it right}.
\label{fig:lofarnustarphoton}}
\end{figure*}

\begin{figure*}
\centering
\includegraphics[width=170mm]{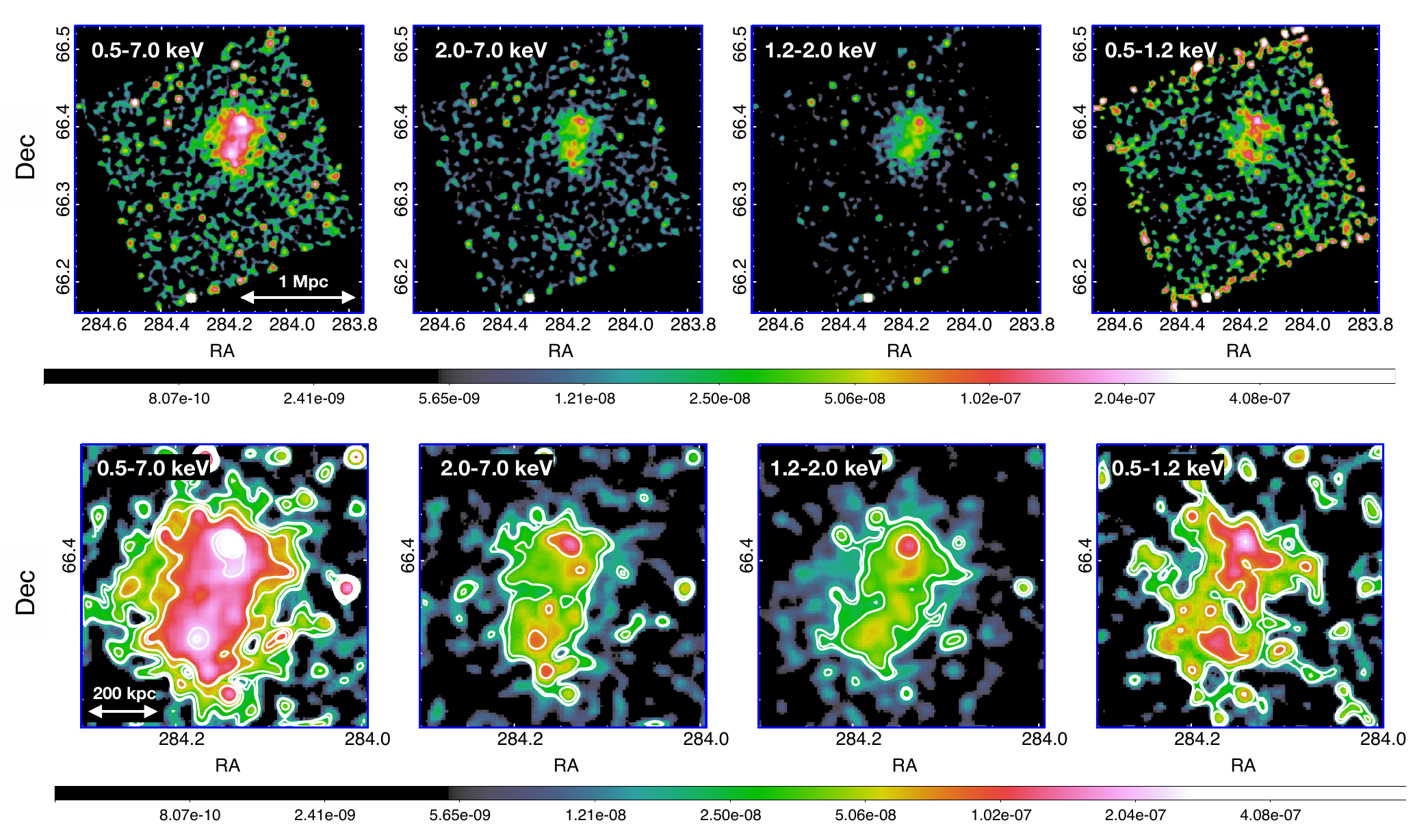}
\caption{Background-subtracted, exposure-corrected, \chandra~flux images of \zwcl~in broad, soft, medium and hard bands smoothed with a Gaussian of r=5 pixels (1 pixel subtending 0.492$\arcsec$). Zoomed in images focusing on the BCGs are presented on the lower panels with \chandra~X-ray contours.
\label{fig:chandraphoton}}
\end{figure*}

We modeled the \nustar~background using a set of IDL routines called {\tt nuskybgd} that has become a standard for background assessment of extended sources in the last decade. The treatment of the background and its components are explained in detail in \citet{wik14}. We applied {\tt nuskybgd} to the spectra extracted from cluster emission free regions at each detector but still accounting for a residual ICM emission with a single temperature model. The joint fit of the background models as well as the contribution from the residual ICM emission for each FPM are shown in Figure~\ref{fig:nuskybgd}. This global background model was used to create background images, which were then subtracted from the images and were corrected by the corresponding exposure maps. Background subtracted, exposure corrected images in the 3.0~--~8.0 keV and 8.0~--~15.0 keV bands are presented in the left and middle panels of Figure~\ref{fig:lofarnustarphoton}. 

\subsection{\chandra}

\chandra~observed \zwcl~in 2018 with Chandra Advanced CCD Imaging Spectrometer Imaging (ACIS-I) array with Observation ID~:~19752 for 42.4 ks of exposure time. For data reduction and spectral extraction we utilized \chandra~Interactive Analysis of Observations (CIAO) version 4.14 package with CALDB 4.9.8. In order to create level-2 files, we used the standard CIAO script {\tt chandra\_repro}. Then we cleaned the data from solar flares using {\tt lc\_clean}. 

{\tt wavdetect} algorithm was used to detect point sources followed by a visual inspection. We used {\tt acis\_bkgrnd\_lookup} script to obtain the blank-sky background file from the CALDB that matched the observation epoch. These blank-sky background event files were used to obtain the flux images. Background subtracted, exposure corrected images in various energy bands are presented in Figure~\ref{fig:chandraphoton}.

\section{Data Analysis and Results} \label{sec:analysis}

\subsection{Imaging Analysis}
In order to search for substructures via surface brightness variations in the cluster, we applied Gaussian Gradient Magnitude (GGM) filter \citep{sanders16} to the background-subtracted, exposure corrected \chandra~photon image in the broad (0.5~--~7.0 keV) band. The GGM filter calculates the gradient of an image assuming Gaussian derivatives with a width of $\sigma$ (pixels). This method captures gradients at different scales depending on the selection of the width, i.e., small $\sigma$ values are used at the central regions where there are many counts, and large $\sigma$ values are better at capturing gradients at cluster outskirts that have lower photon counts. To create the GGM filtered image, the image itself is convolved with the gradient of a 1D Gaussian function for two axes, then these two resulting images are combined for the 2D gradient image.

\begin{figure}
\centering
\includegraphics[width=86mm]{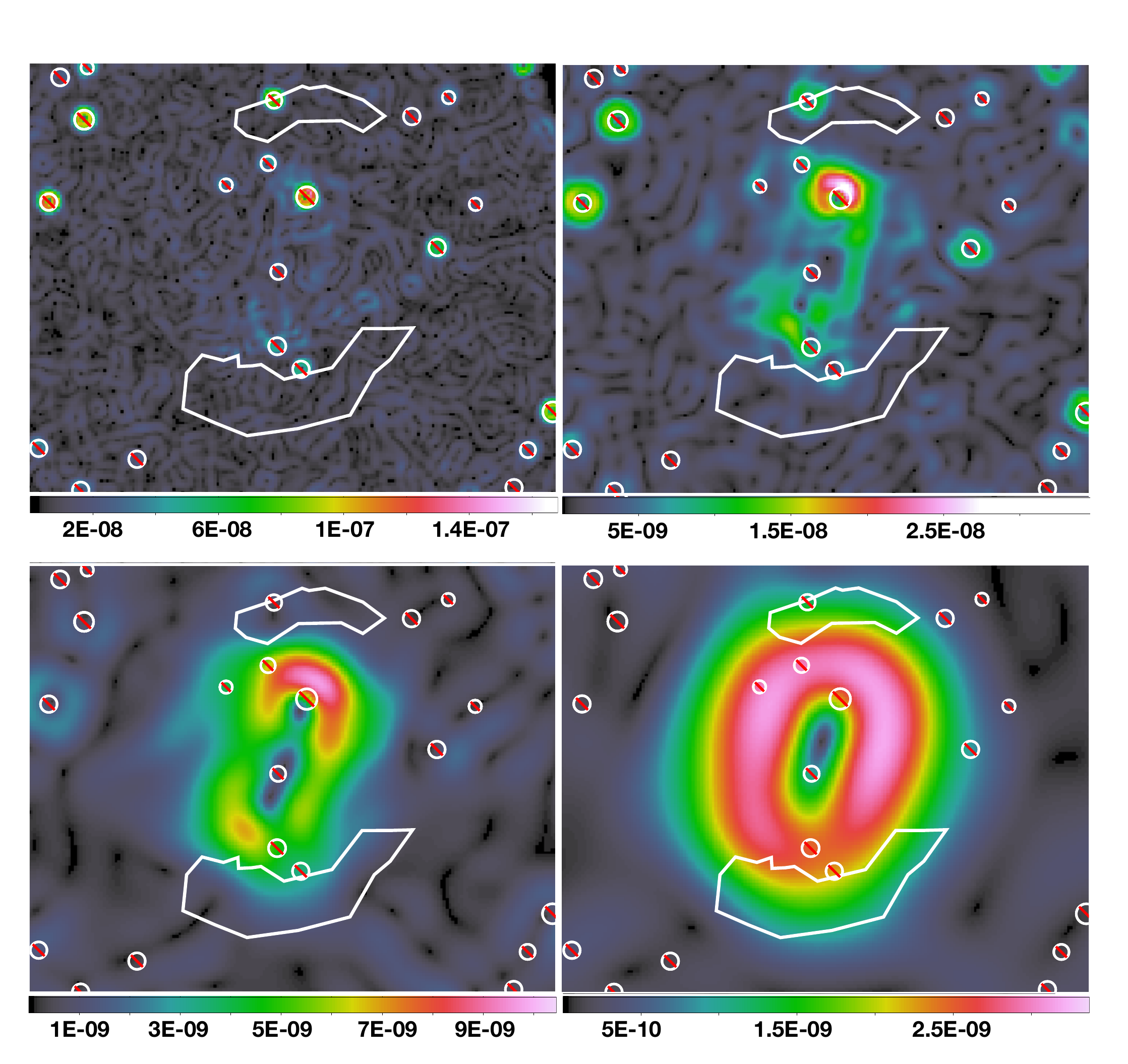}
\caption{\chandra~GGM filtered background subtracted, exposure corrected images in the 0.5~--~7.0~keV band with $\sigma$~=~2, 4, 8, and 16 pixels from; upper left, upper right, lower left and lower right panels, respectively. White polygons roughly trace the radio relics seen by LOFAR. The scales emphasize the regions with largest SB gradients and does not accurately reflect the magnitude of the jumps \citep{sanders16}. Large gradients such as edges, are indicated with lighter colors and flatter SB areas are darker colors.}
\label{fig:GGM}
\end{figure}

In Figure~\ref{fig:GGM} we present GGM filtered images with $\sigma$~=~2, 4, 8, and 16 pixels, where we overlaid LOFAR contours of the radio relics. Due to the lack of deeper data, GGM $\sigma$~=~1 image is mainly dominated by the noise, therefore we excluded it from this figure. This also can be said for $\sigma$~=~2, however, it is somewhat important to show where gradients begin to appear. 

A strong, asymmetrical gradient connecting the two subclusters appears with $\sigma$~=~4. In the same image, we see that northern subcluster is more pronounced than the southern subcluster. GGM filter with $\sigma$~=~8 hints at a surface brightness edge within the southern relic polygon region, where a similar edge seems to reach the inner part of northern relic. At $\sigma$~=~16, GGM shows a relatively symmetrical gradient distribution, yet sharper edges are present in the northern regions.

Since the main motivations of this work are to search for any shock features coinciding with the relics, as well as any radio relic~-~IC connection, we defined the regions of interest as explained in the following section.

\subsection{Spectral Analysis}\label{sec:specanalysis}
\subsubsection{Temperature distribution within the ICM}

\begin{figure}
\centering
\includegraphics[width=80mm]{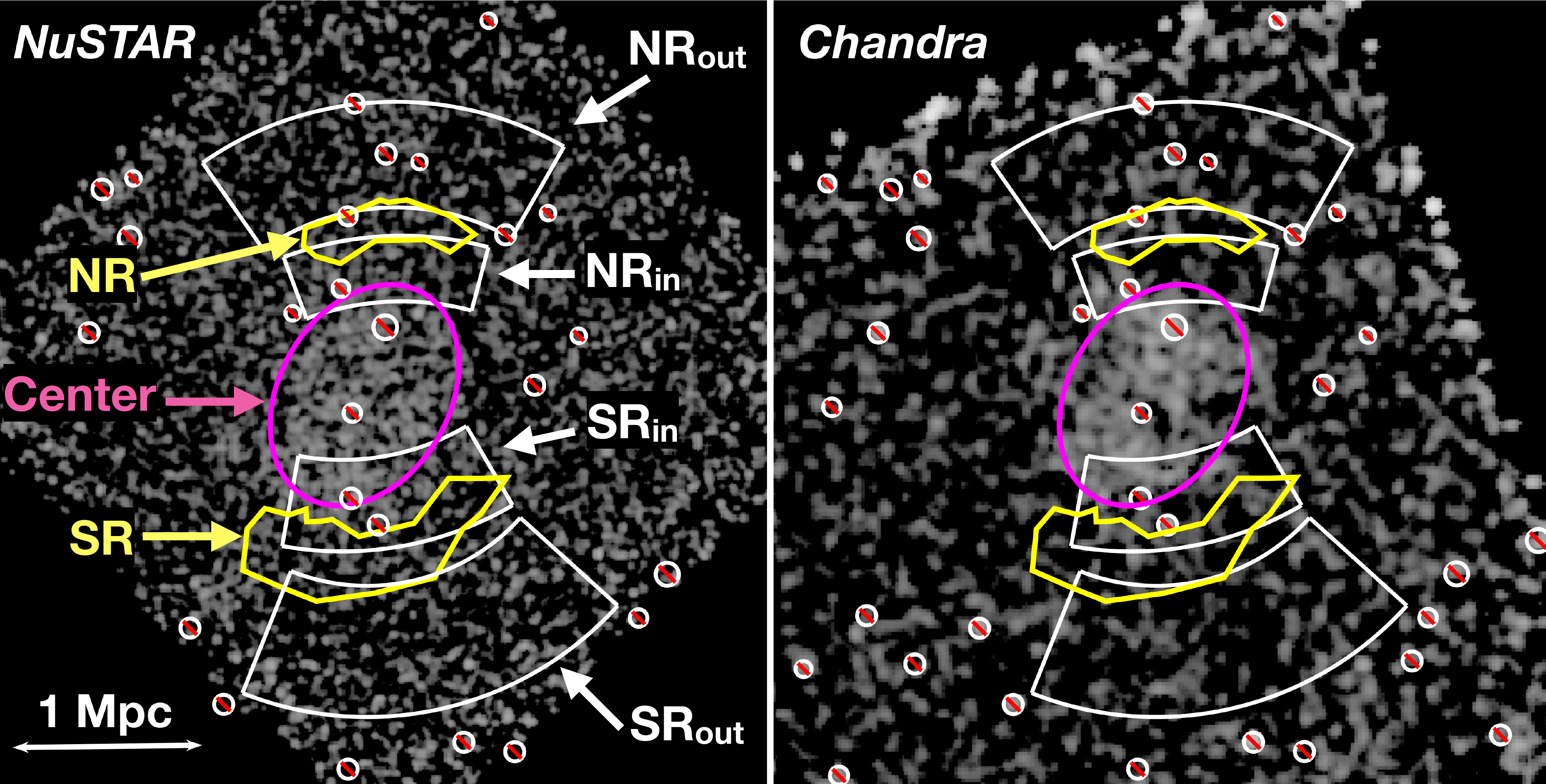}
\caption{Regions of interest overlaid on \nustar~({\it left panel}) and \chandra~images ({\it left panel}).
\label{fig:regions}}
\end{figure}

In line with the motivation of this work, we selected seven regions of interest. ``Center" region, defined as an ellipse, is selected to assess the subcluster features of the system enclosing the extended emission pronounced in photon images (Fig~\ref{fig:lofarnustarphoton} and Fig~\ref{fig:chandraphoton}). We then defined two polygons tracing the northern (NR) and southern (SR) relics from the LOFAR data, to study the thermal and non-thermal emission contributions. Four sectors are selected to measure the temperature of the inner and outer regions of the northern and southern relics, i.e.  NR$_{in}$, NR$_{out}$, SR$_{in}$ and SR$_{out}$. These regions are shown in Figure~\ref{fig:regions}.

\nustar~spectra were extracted using {\tt nuproducts}, followed by background extraction using {\tt nuskybgd} that creates spectra from the model defined across the FOV as defined in Section~\ref{sec:reduction}.
For analyzing the \chandra~data, we extracted the source spectra using {\tt specextract} task, which also creates the weighted RMFs and ARFs. A local background extraction approach from the same observation was chosen given the redshift of the cluster. With this {\tt specextract} task, we simultaneously extracted a background spectrum from a region with $r=3\arcmin$ in the field of view where no cluster emission is expected. 

\begin{figure}
\centering
\includegraphics[width=89mm]{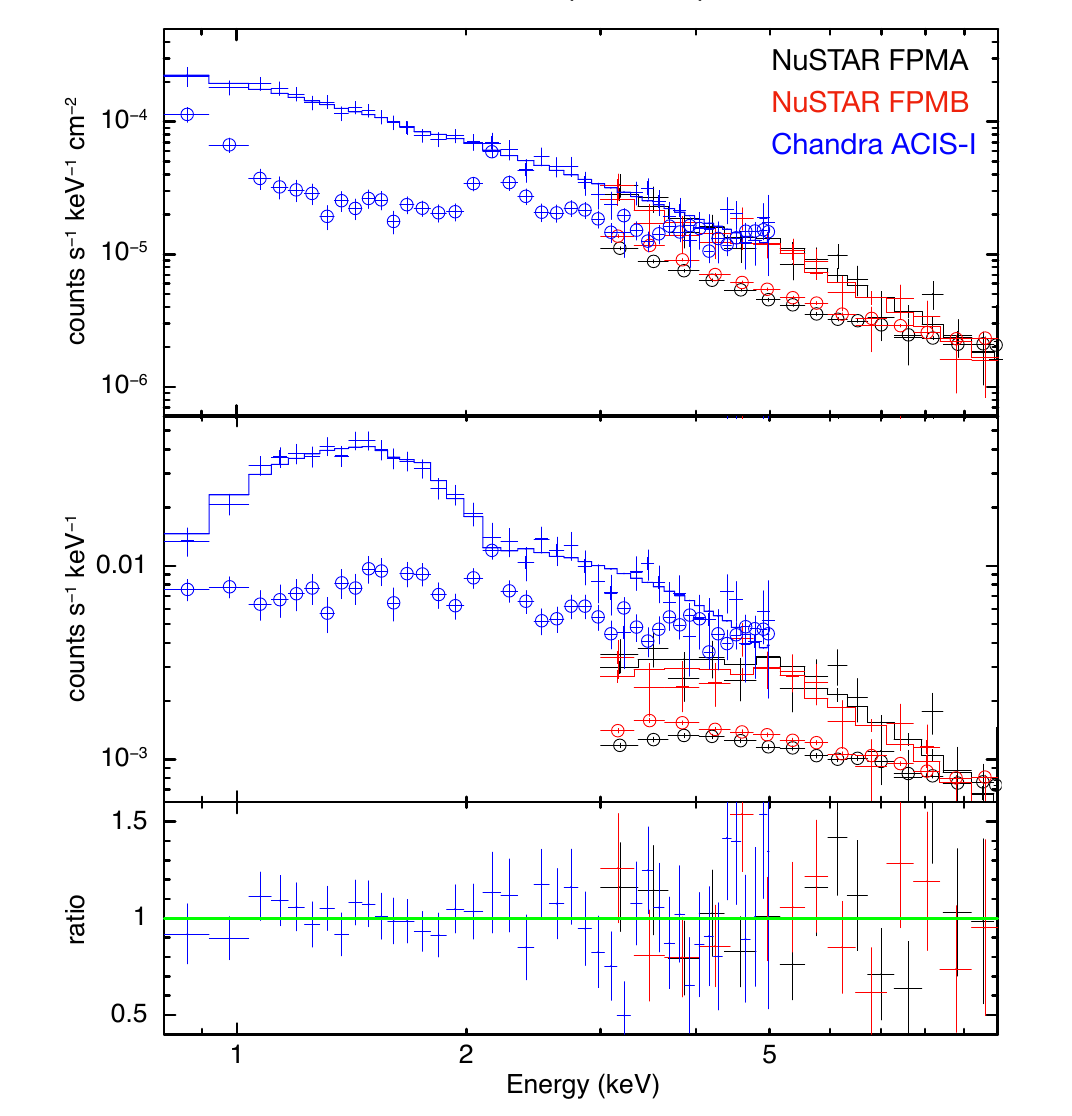}
\caption{Central region spectra showing \nustar~(black and red) and \chandra~(blue) joint fits using a single temperature model.
The corresponding blue, black, and red circles below the source spectra indicate the background level for each instrument.
\label{fig:centerspectra}}
\end{figure}

For spectral fitting, we use {\tt XSPEC}\footnote{\url{https://heasarc.gsfc.nasa.gov/docs/xanadu/xspec/}}. We apply maximum likelihood-based statistic (hereafter, C-stat) that is appropriate for Poisson data as proposed by \citet{cash79}. Photon counts used in spectral analysis for both \nustar~and \chandra~spectra are grouped to have at least 3 counts in each bin. We could only use 3.0~--~10.0~keV band of \nustar's bandpass due to low S/N at high energies where background is dominant, and used 0.5~--~7.0~keV band for \chandra. While the hard energy information beyond 10.0~keV is extremely important to differentiate the thermal vs. non-thermal emission components, slightly lower global temperature of this cluster with respect to a typical violent merger encourages the spectral analysis with the data at hand even in the 3.0~--~10.0~keV band.

\movetabledown=50mm
\begin{rotatetable*}
\begin{deluxetable*}{l|ccc|ccc|ccccc}
\tabletypesize{\scriptsize}
\tablewidth{0pt} 
\tablecaption{Parameter values obtained from individual and joint fits to the \nustar~and \chandra~spectra using single temperature model from the regions show in Figure~\ref{fig:regions} and plotted in Figure~\ref{fig:spectralfitNuCH}. \textit{N$_{H}$} is kept fixed at 4.45 $\times$ 10$^{20}$~cm$^{-2}$. The {\tt apec} normalization ({\it norm}) is given in $\frac{10^{-14}}{4\pi \left [ D_A(1+z) \right ]^{2}}\int n_{e}n_{H}dV$. 
%(10$^{-3}$~cm$^{-5}$) where {\tt powerlaw} normalization ($\kappa$) is photons~keV$^{-1}$~cm$^{-2}$~s$^{-1}$ at 1 keV (10$^{-3}$).
\label{tab:nuchfit}}
\tablehead{%\\[-0.95em]
\colhead{} & \multicolumn{3}{c}{\nustar~Fit} & \multicolumn{3}{c}{\chandra~Fit} & \multicolumn{5}{c}{Joint Fit} %\\[-0.95em]
\\
\hline
\colhead{} &  \colhead{kT} & \colhead{norm} & \colhead{} & \colhead{kT} & \colhead{norm} & \colhead{} & \colhead{kT} & \colhead{norm$_{FPMA}$} & \colhead{norm$_{FPMB}$} & \colhead{norm$_{Chandra}$} &\colhead{} \\[-0.95em]
\colhead{Region} &  \colhead{(keV)} & \colhead{(10$^{-3}$~cm$^{-5}$)} & \colhead{C/$\nu$} & \colhead{(keV)} & \colhead{(10$^{-3}$~cm$^{-5}$)} & \colhead{C/$\nu$} & \colhead{(keV)} &  \colhead{(10$^{-3}$~cm$^{-5}$)} &\colhead{(10$^{-3}$~cm$^{-5}$)} &\colhead{(10$^{-3}$~cm$^{-5}$)} &\colhead{C/$\nu$}}
\startdata
\\[-0.95em]
NR$_{out}$ & 8.51$^{+34.67}_{-5.54}$& 0.198$^{+0.374}_{-0.074}$  & 144.38/130 & ... & $<$0.029 &228.29/220 & 8.82$^{+24.61}_{-5.25}$ & 
0.121$^{+0.183}_{-0.050}$ & 0.284$^{+0.401}_{-0.111}$ & 0.002$^{+0.021}_{-0.002}$ &
367.59/348\\
\\[-0.5em]
NR& 1.95$^{+1.65}_{-0.93}$ & 0.553$^{+3.879}_{-0.383}$  & 34.10/33 & 1.12$^{+0.58}_{-0.71}$ &0.054$^{+0.205}_{-0.018}$ &60.98/71 & 1.34$^{+1.03}_{-0.31}$ & 
0.742$^{+1.575}_{-0.596}$ & 2.445$^{+4.310}_{-1.886}$ & 0.047$^{+0.019}_{-0.015}$ &
89.54/102\\
\\[-0.5em]
NR$_{in}$ & 4.37$^{+1.74}_{-1.04}$&0.371$^{+0.179}_{-0.117}$  & 88.09/77 & 3.80$^{+2.16}_{-1.00}$&0.157$^{+0.023}_{-0.022}$&127.73/156 & 4.07$^{+1.03}_{-0.31}$ & 
0.260$^{+0.114}_{-0.081}$ & 0.553$^{+0.216}_{-0.157}$ & 0.154$^{+0.019}_{-0.017}$ &
205.32/231\\
\\[-0.5em]
Center & 6.40$^{+0.85}_{-0.87}$ & 1.349$^{+0.205}_{-0.141}$ & 203.60/231 & 6.21$^{+0.79}_{-0.67}$ & 1.294$^{+0.042}_{-0.039}$ & 230.93/283 & 6.30$^{+0.57}_{-0.55}$ & 
1.368$^{+0.148}_{-0.125}$ & 1.252$^{+0.143}_{-0.121}$ & 1.291$^{+0.038}_{-0.036}$ &
434.56/512\\
\\[-0.5em]
SR$_{in}$ & 4.77$^{+0.98}_{-0.67}$&0.986$^{+0.217}_{-0.186}$  & 153.61/160 & 4.72$^{+1.28}_{-0.93}$ &0.409$^{+0.034}_{-0.029}$ &187.14/229 & 4.76$^{+0.73}_{-0.56}$ & 
1.033$^{+0.198}_{-0.169}$ & 0.847$^{+0.176}_{-0.149}$ & 0.408$^{+0.027}_{-0.026}$ &
340.13/387\\
\\[-0.5em]
SR & 3.05 $^{+2.73}_{-1.23}$&0.401$^{+0.738}_{-0.230}$  & 104.40/96 & 3.79$^{+4.85}_{-1.63}$& 0.096$^{+0.027}_{-0.020}$& 141.65/169 & 3.50$^{+2.04}_{-1.15}$  &
0.364$^{+0.378}_{-0.170}$ & 0.251$^{+0.315}_{-0.139}$ & 0.098$^{+0.022}_{-0.019}$ &
245.78/263\\
\\[-0.5em]
SR$_{out}$ & 2.35$^{+2.28}_{-1.34}$&1.851$^{+21.026}_{-1.220}$  & 209.39/221 & $>$6.49 &0.055$^{+0.033}_{-0.035}$& 268.36/270 & 3.38$^{+4.34}_{-2.07}$ & 
0.739$^{+7.531}_{-0.486}$ & 1.255$^{+11.339}_{-0.780}$ & 0.035$^{+0.031}_{-0.030}$ &
476.64/489\\
\\[-0.5em]
\enddata
\end{deluxetable*}
\end{rotatetable*}

To begin with, we fitted the spectra of \nustar~and \chandra~individually. The initial assessment of the temperatures were achieved using a single temperature {\tt apec} model whose redshift and metal abundance parameters were fixed at z~=~0.304, and \textit{Z$_{1}$}~=~0.3~{\it Z$_{\odot}$}, respectively, due to low photon statistics at hand. This model was then convolved with the foreground hydrogen column density modeled by {\tt phabs}, where the  N$_{H}$ parameter was fixed at 4.45$\times$10$^{20}$~cm$^{-2}$ (based on Leiden/Argentine/Bonn Galactic HI survey \citep{kalberla05}). We used the abundance table of Wilms \citep{wilms00}. This single temperature model assumes an isothermal ICM. 

We then fit the \nustar~and \chandra~spectra jointly. During this procedure, the {\tt apec} temperature, abundance and redshift parameters were linked within the instruments. We accounted for the cross-calibration differences of detectors with the {\tt constant} model, where the normalization of the models of different instruments was tied. The joint-fit model hence becomes; {\tt constant}~$\times$~{\tt phabs}~$\times$~{\tt apec}. For the Center region, where we have the highest S/N, we could allow the cross calibration constant parameter be free. Namely, one instrument constant is fixed to 1, whereas the other instrument constant parameters are left free to vary. The cross-calibration constants of \chandra~(0.94) and \nustar~FPMs are within 6\%, and FPMA constant (fixed to 1) and FPMB (0.92) constant are within 8\% of each other. The temperature resulting from this fit is 6.30$^{+0.37}_{-0.35}$~keV with C-stat/d.o.f. (C/$\nu$) 434.56/514. The \nustar~and \chandra~joint spectral fit for the Center region using {\tt constant}~$\times$~{\tt phabs}~$\times$~{\tt apec} is shown in Figure~\ref{fig:centerspectra} as an example of our fitting procedure. 

As we move away from the central region to the cluster outskirts, where the S/N decreases and the background begins to dominate, the cross-correlation constants become incongruent, pointing to discrepancies in \nustar~and \chandra~fluxes and/or systematics in the background subtraction, resulting in unconstrained and unrealistic temperatures. Hence, we removed the cross-correlation constant parameter and let normalizations of all instruments be free and kept temperature, abundance and redshift parameters still tied amongst the instruments. Only then the joint-fit temperature values reflected what was obtained from the individual spectral fits. 

We note that although this approach gives new information on the temperature parameter, it limits any interpretations of the densities that are obtained from the normalization values of these joint fits since all normalization are free and the which instrument represent the correct flux is not obvious.

\begin{figure}
\centering
\includegraphics[width=80mm]{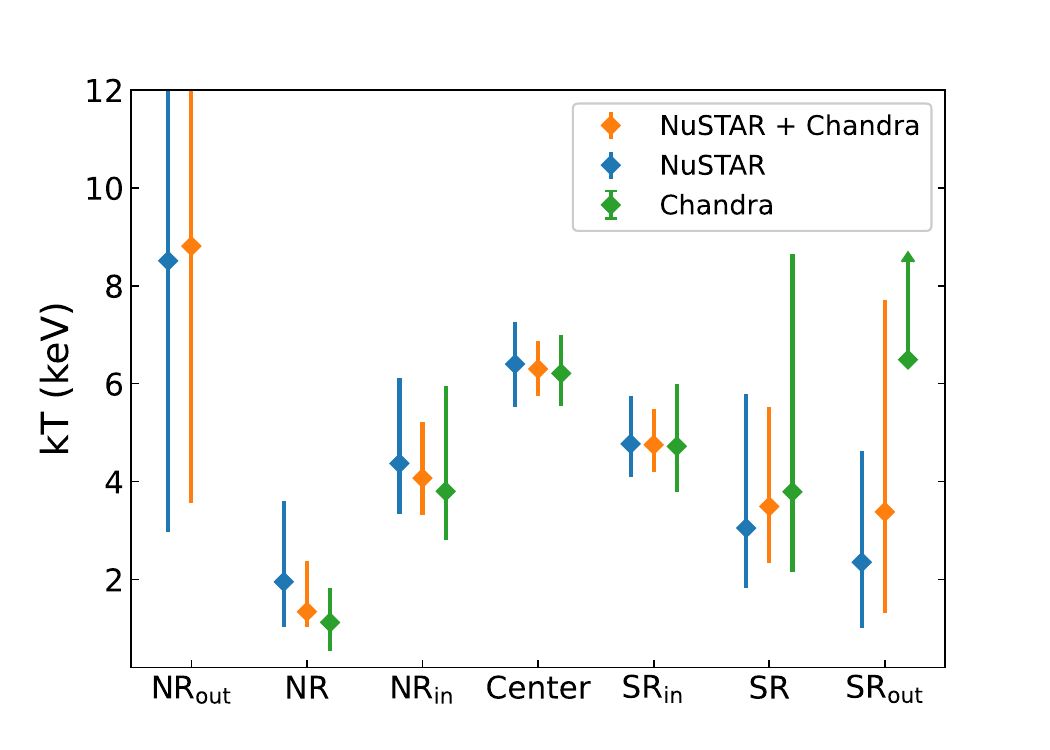}
\caption{Best-fit temperatures in each region. \chandra~temperature is unconstrained at NR$_{out}$ hence is not shown. \nustar~temperature upper limit of $\sim$42~keV (Table~\ref{tab:nuchfit}) in addition to \nustar~and \chandra~joint fit temperature upper limit of $\sim$58~keV are cropped from the top to enable better visualization of other error bars.
\label{fig:spectralfitNuCH}}
\end{figure}

We present the fit parameter values of individual fits as well as the joint fits with free instrument normalizations in Table~\ref{tab:nuchfit}. The temperature results are plotted in Figure~\ref{fig:spectralfitNuCH} with regions ordered from North to South to provide a visualization of a cluster radial profile.

We also fitted the Center region with two other models; a two temperature thermal ICM (2T; {\tt constant}~$\times$~{\tt phabs}~$\times$~[{\tt apec}~+~{\tt apec}]) and a thermal plus non-thermal model (1T+IC; {\tt constant}~$\times$~{\tt phabs}~$\times$~[{\tt apec}~+~{\tt powerlaw}]). We let the cross correlation constants free for these procedures.

The resulting temperature values from 2T model temperatures are 6.75~keV and 1.45~keV, both unconstrained. The statistics yield C/$\nu$ = 434.01/510, and {\tt constant} parameters are 1.00 (fixed), 0.91 and 0.94 for \nustar~FPMA, \nustar~FPMB, and \chandra, respectively. Only when these constant parameters are fixed to their best-fit values, we were able to find a lower limit to the high temperature component of 6.21~keV.

The 1T+IC model results in a {\tt apec} temperature value of 6.45$^{+0.81}_{-1.57}$~keV and a {\tt powerlaw} photon index ($\Gamma$) value of 2.41 (unconstrained) with C/$\nu$~=~434.43/510 and same {\tt constant} values found for 2T model. When we fixed the {\tt constant} parameter, $\Gamma$ was still unconstrained and the temperature and the corresponding errors become 6.45$^{+0.67}_{-1.47}$~keV.

Finally, we removed the {\tt constant} component from the 2T and 1T+IC models, leaving all instrument normalizations free and we present the results in Table~\ref{tab:compare}.

To quantify the systematics of the background characterization, we applied a generous $\pm$15\% change to the total background of the central region spectrum of \chandra~and \nustar~spectral fits, and reran the individual fits. The temperature value changes were within 1.1 $\sigma$ for \chandra~and 0.9 $\sigma$ of the statistical errors on the temperature of that region. Given that the background level is unlikely to be off by this amount, we conclude that the measurements are not particularly sensitive to it.

\subsubsection{Relics}

For the relic regions, we also investigated the possible non-thermal emission components. This was achieved in the similar manner that was applied to the Center region, leaving the cross normalizations of instruments untied and no additional {\tt constant} parameter. We applied a two temperature thermal ICM (2T), and a thermal plus non-thermal model (1T+IC) to the joint \nustar~and \chandra~spectra. 2T model does not only test the possibility of multi-temperature structure, but also is a good judgement of the validity of the 1T+IC model. Mainly, if the statistics favour 1T+IC and 2T models equally well, then the conclusion of a dominant IC emission within the ICM requires further investigations. When the low and high temperature components of the 2T model have close values, this is a good confirmation that the ICM can be described well by a single temperature model. On the other hand, 2T model also emphasizes idiosyncrasies of the data (low S/N, background assessment issues) when it results in unconstrained or unphysical temperature values. Therefore, it is a good practice to apply 2T model to the data prior to making conclusions about the contribution of non-thermal emission.

Although thermal emission should be present at the relic sites we also fit each relic region spectra jointly with \nustar~and \chandra~spectra using a single IC model. This approach provides a higher limit to the detection significance with the assumption that ICM emission is entirely dominated by the non-thermal IC emission. 

\textbf{Northern Relic:} 2T model fit to the joint \nustar~and \chandra~NR region spectra results in a lower temperature component of {\it kT}~=~0.40$^{+0.29}_{-0.27}$~keV and a higher temperature component of 3.25~keV, with a lower limit of 1.55~keV. The resulting statistics is C/$\nu$~=~86.72/98. 

When we reintroduced the {\tt constant} parameter with values fixed to Center region values (hence tied the normalizations), one of the temperatures hit the 64~keV hard limit, and the other temperature became {\it kT}~=~1.11$^{+0.54}_{-0.31}$~keV with C/$\nu$~=~88.28/100. For comparison, 1T model fit to the joint \nustar~and \chandra~spectra of the same region is C/$\nu$~=~89.54/102 ({\tt constant} parameter.) 

1T+IC implementation of this region results in unconstrained temperature and photon index values. When we repeat this fit by fixing the photon index to $\Gamma$~=~1.87 obtained from $\alpha_{inj}$ reported by \citet{jones21} ($\Gamma$~=~$\alpha$+1), we get {\it kT}~=~1.02$^{+0.39}_{-0.61}$~keV with C/$\nu$~=~87.27/99. This is based on the assumption that the synchrotron emitting electron population is also responsible for the IC emission.

Single {\tt powerlaw} fit to the NR results in a $\Gamma$~=~4.31$^{+0.91}_{-0.79}$ with C/$\nu$~=~86.14/102. This fit implies a significance of $\sigma$~=~2.1.
We also repeated this fit by fixing the $\Gamma$ to 1.87 implied by the radio spectral index found by \citet{jones21}. With this approach, the detection significance becomes; $\sigma$~=~1.9 with fit statistics C/$\nu$~=~96.21/103

\textbf{Southern Relic:} When applied these same procedure steps to the southern relic. 2T model results in a higher and lower temperature components of {\it kT}~=~9.84~keV(unconstrained) and {\it kT}~=3.40$^{+7.23}_{-2.45}$~keV, respectively, both unconstrained. The resulting statistics for this fit is C/$\nu$~=~245.79/261, where 1T model fit to the joint \nustar~and \chandra~spectra of the same region is C/$\nu$~=~245.78/263. Again, we introduced the {\tt constant} parameter values fixed to the Center values, and the temperatures for both models become {\it kT}~=~8~keV (unconstrained) with C/$\nu$~=~250.46/263.

1T+IC modeling of this region results in an extremely volatile photon and temperature values. We repeated the same procedure applied to the northern relic 1T+IC fit by fixing the $\Gamma$ to 1.97 using $\alpha_{inj}$ value \citet{jones21}, we find {\it kT}~=~2.30$^{+1.05}_{-1.50}$~keV with C/$\nu$~=~245.15/260.

When we applied the single {\tt powerlaw} model, we find a photon index value of $\Gamma$~=~2.49$^{+0.51}_{-0.44}$ with C/$\nu$~=~247.06/263. With the aforementioned assumption of the origin of the emission and model use, we put of detection limit of at least 5$\sigma$. When we use $\Gamma$~=~1.97, the significance becomes 6$\sigma$.

In Table~\ref{tab:compare}, we provide a comparison table of the findings from 1T, 2T, and 1T+IC model implementations on the Center, NR and SR region spectra and additional IC model results to the NR and SR regions.

\begin{deluxetable*}{lcccc}
\tabletypesize{\scriptsize}
\tablewidth{0pt}
\tablecaption{This table represents the results of 1T, 2T, and 1T+IC model fits of the joint \nustar~and \chandra~} regions Center, NR and SR, with the additional IC model results of NR and SR spectral fits. Abundances are fixed to \textit{Z$_{1}$}~=~0.3~{\it Z$_{\odot}$} and instrument normalizations are left free to vary.

\label{tab:compare}
\tablehead{
\colhead{Region} & \colhead{Model} & \colhead{$kT$} & \colhead{$kT$ or $\Gamma$} & \colhead{C-stat/d.o.f.} \\[-0.5em]
 &  & \colhead{(keV)}  & \colhead{(keV or \#)} & 
}
\startdata
Center & 1T & $6.30^{+0.57}_{-0.55}$  & {---} & 434.56/512  \\
& 2T & $6.52^{+0.68}_{-0.60}$  & $0.29^{+0.27}_{-0.22}$ & 432.23/508 \\
& 1T+IC & $6.23^{+0.81}_{-0.67}$  & $2.41$ & 433.84/508 \\
\hline
\\[-0.95em]
NR & 1T & $1.35^{+1.03}_{-0.32}$  & {---} & 89.54/102 \\
& 2T & $3.25^{+...}_{-1.70}$ & $0.40^{+0.29}_{-0.27}$ & 86.72/98 \\
& 1T+IC & $1.02^{+0.39}_{-0.61}$ & 1.87 (fixed) & 87.27/99 \\
& IC & {---} & $4.31^{+0.91}_{-0.79}$ (1.87 [fixed]) & 86.14/102 (97.47/103)  \\
\hline
\\[-0.95em]
SR & 1T & $3.50^{+2.04}_{-1.15}$  & {---} & 245.78/263  \\
& 2T & $9.84^{...}_{-...}$  & $3.40^{+7.23}_{-2.45}$ & 245.12/259 \\
& 1T+IC & $2.02^{+1.57}_{-1.27}$ & 1.97 (fixed) & 245.15/260 \\
& IC & {---} &  $2.49^{+0.51}_{-0.44}$ (1.97 [fixed]) & 247.06/263 (248.44/264) \\
\\[-0.95em]
\enddata
\end{deluxetable*}

\subsection{Cross-talk Analysis}\label{sec:cross-talk}

Even single bright sources in the \nustar~FOV cause a scatter of photons due to the large ($\sim$1$\arcmin$ Half Power Diameter, $\sim$18$\arcsec$ Full Width at Half Maximum), slightly energy-dependent point spread function (PSF). This results in a cross-contamination, namely {\it cross-talk}, of multiple emission features in the regions of interest, although the emission may not purely originate from the selected region. Standard \nustar~pipeline, {\tt nuproducts}, produces ARFs for point or diffuse sources inside the user defined extraction regions. However, it does not account for the ARFs for other sources whose emission originates outside these extraction regions that contaminate the spectra of those regions, which will be referred to as cross-ARFs.

A set of IDL routines created to account for this contamination that divorces the contamination and source emission is {\tt nucrossarf}\footnote{\url{https://github.com/danielrwik/nucrossarf}}. This method is explained, applied and tested in detail by \citet{tumer23}.

\begin{figure}
\centering
\includegraphics[width=80mm]{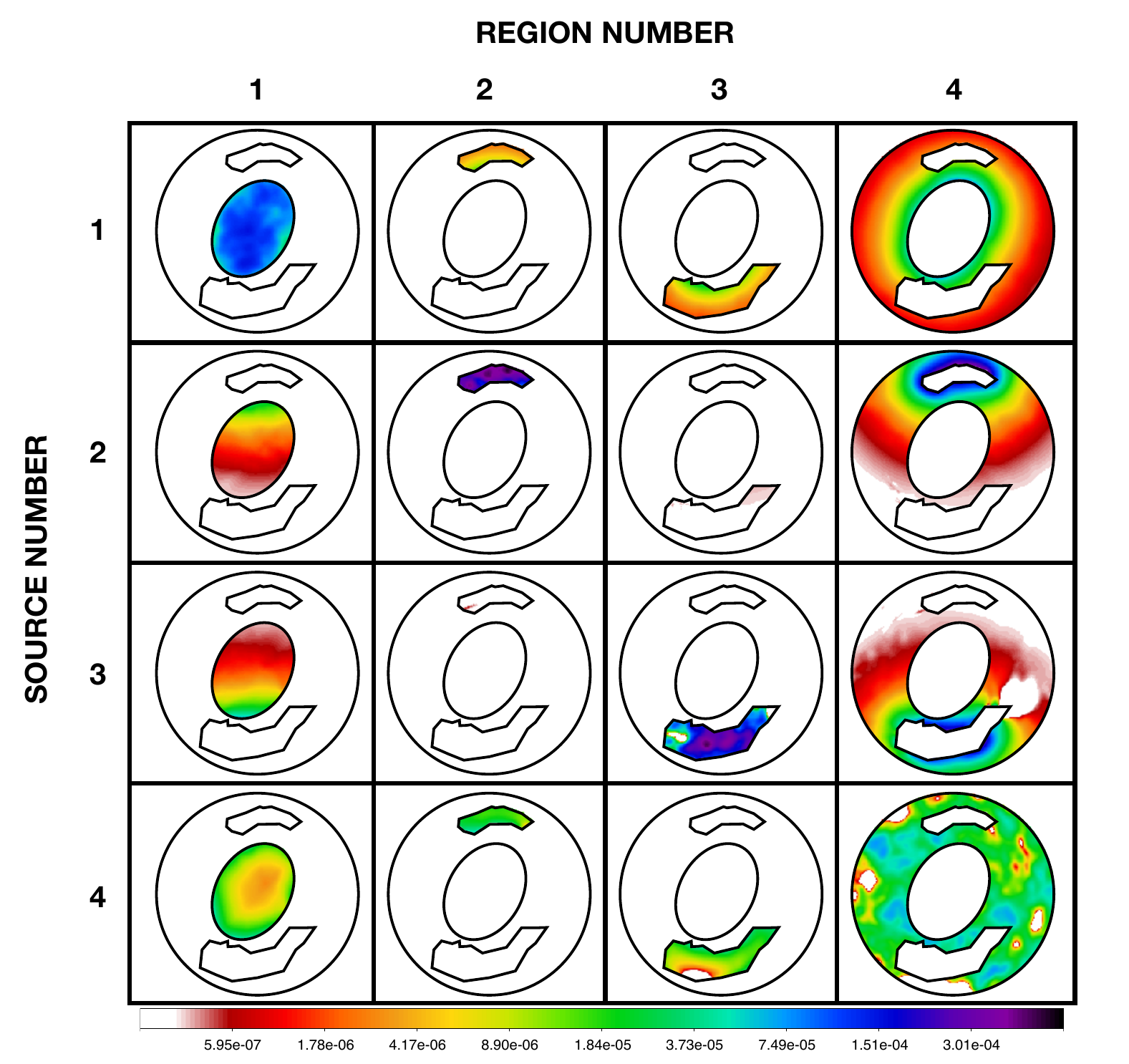}
\caption{{\tt nucrossarf} analysis images showing the relative emission scattered by the PSF from one region into the others. Row numbers on the left indicate the region from which source emission originates, and the column numbers on the top indicate the region number into which the source photons leak. The diagonal images show the actual source image for each of the four regions.
The scale of the logarithmic color bar has arbitrary units.
\label{fig:crossarfimage}}
\end{figure}

\begin{figure}
\centering
\includegraphics[width=80mm]{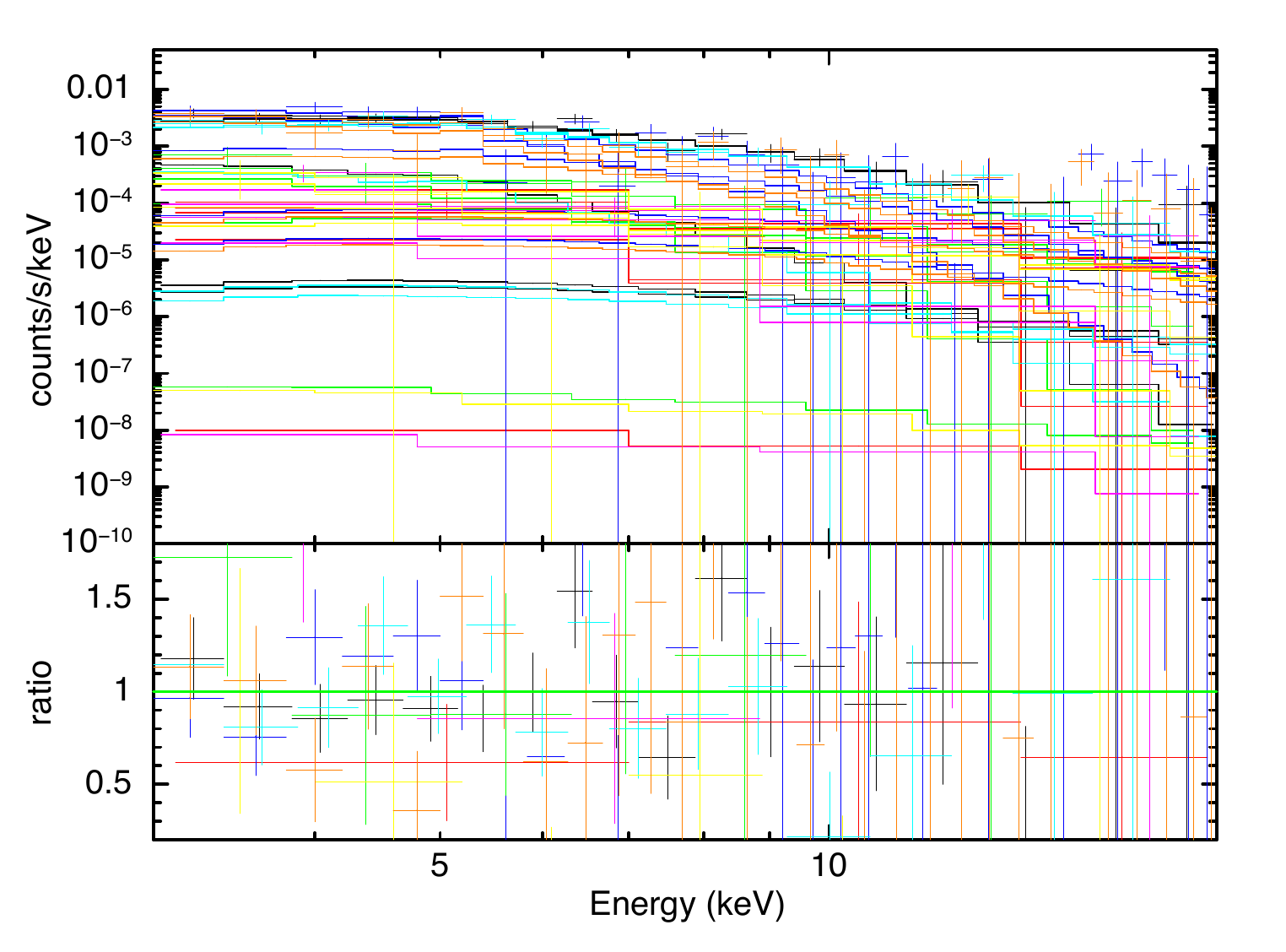}
\caption{Joint cross-talk spectral fit from all regions shown in Fig~\ref{fig:crossarfimage}, comprising of four source spectra, four source models and the 3$\times$4 cross-talk models.
\label{fig:crossarffit}}
\end{figure}

Given the lack of good photon statistics, we were forced to simply our regions of interest for the cross-ARF analysis to increase the photon statistic. We kept the Center, NR, and SR regions to assess the central temperature more accurately, as well as to constrain the relic emission better to study the implications of a possible IC emission. We also selected a circular region encircling but excluding the aforementioned regions, which we will refer to as; ``Circle". This way, the contribution from both the dominant Center emission and the cooler gas beyond the central region is separated.

While running the {\tt nucrossarf} an order is assigned to each region. In our treatment, the order from 1-4 is designated to Center, NR, SR, and the Circle regions, respectively. To assess the cross-talk, a model is assigned to each source distribution. Whereas we expect a dominant thermal ICM emission ({\tt apec}) from the Center and Circle regions, we applied a {\tt powerlaw} model to represent a non-thermal (IC) emission coming from the relics to assess the statistical significance assuming the relic emission is purely from the IC. Although, {\tt nucrossarf} allows assigning multiple models to each region emission for a more accurate representation of complex structures, the lack of good photon statistics of the current data dictates a simplistic approach. Namely; we expect a considerable amount of photons to leak inside the relic regions from the more dominant thermal ICM of the Center region, hence this will decrease the number of photons originating from the relic regions. Since the initial fits of the relic regions already resulted in large uncertainties, especially in the SR region (Table~\ref{tab:nuchfit}), this will worsen with fewer photons once the cross-talk is applied.

{\tt apec} model redshift and metal abundance values were fixed to \textit{z}~=~0.304, and \textit{Z$_{1}$}~=~0.3~{\it Z$_{\odot}$}, respectively as in the original fits without the cross-talk analysis.

The visual 4~$\times$~4 matrix given in Figure~\ref{fig:crossarfimage} presents the mapping of the contribution from the local photons within a region and the cross-talk, where the selected regions are overlaid. In this M~$\times$~N matrix, M represents the local source PSF and N represents the region number, i.e., the box at 2~$\times$~4 shows the PSF leakage from the source situated at region 2 into Region 4. In other words, i~$\times$~i boxes represent the PSF distribution confined with the source region, whereas i~$\times$~j shows the extend of that PSF with other regions of interest.

The spectral distribution of the source emission, the individual source ARFs and the cross-ARFs are shown in in Figure~\ref{fig:crossarffit}. This plot is the joint fit result of all 4 source distributions with their respective models, as well as the model contributions from other regions. 

The resulting fit gives ICM temperature of {\it kT}~=~6.27~keV with a lower limit of {\it kT}~=~5.98~keV for the Center region, and {\it kT}~=~1.99$^{+0.73}_{-0.28}$ for the Circle region.

With the data at hand, we find that they are consistent with IC-like emission at the 2.3$\sigma$ and 1.1$\sigma$ significance levels for the northern and southern relics, respectively, where $\Gamma$ is fixed at 2 as suggested by the radio spectral index found by \citet{jones21}. When we free the $\Gamma$ parameter, the fit reaches to the lower limit of $\Gamma$=1.
More data are needed to conclusively determine the presence of IC emission in these regions.

\section{Discussion}\label{sec:discussion}

Full band \chandra~photon image and surface brightness contours clearly show X-ray peaks located near the BCGs of the subclusters and bright emission in a region connecting  these halos. \chandra~images do not indicate the inverse-S shape seen by \xmm~ images, which is interpreted as the merger having a non-zero impact parameter by \citet{finner21}. However, GGM filter of the \chandra~image (Figure~\ref{fig:GGM}, upper right panel) clearly shows an strong arc-like gradient to the east, supporting this claim. Moreover, another strong gradient to the north (Figure~\ref{fig:GGM}, upper right and lower left panel) may indicate that the northern halo has stripped material from the southern halo during the first core passage. \nustar~image in the soft band (Figure~\ref{fig:lofarnustarphoton}, left panel) do not reveal any important information, yet the lack of strong emission in the hard band (Figure~\ref{fig:lofarnustarphoton}, middle panel) points to the lack of hot ICM or strong non-thermal emission, yet deeper data is required to make a definite claim.

The spectra extracted from the central region, Center, is best described by a single temperature model indicating a lack of a cool core remnant as well as a robust non-thermal emission.

In addition, to compare our results with the literature \citep{finner21}, we also used a central r~=~3$\arcmin$ region for spectral fitting with \chandra~data to compare \xmm~and \chandra~results. Our temperature results of \chandra~fit agree with \xmm~results within 1$\sigma$ for this same region selection. While \chandra~gives a temperature of {\it kT}~=~4.00$^{+0.43}_{-0.36}$~keV using N$_{H}$~=~9.90$\times$10$^{20}$~cm$^{-2}$, the column density value used by \citet{finner21}, \xmm~temperature is {\it kT}~=~3.7$^{+0.6}_{-0.5}$~keV.

We applied four different models to the NR and SR region spectra to investigate single (1T) and two temperature (2T), thermal plus non-thermal (1T+IC) and pure non-thermal emission by jointly fitting the \nustar~and \chandra~data. Application of 2T model  was aimed at testing the statistical significance of a thermal plus non-thermal model 1T+IC) over a two temperature model rather than searching for a physical multi-temperature plasma structure. 

1T+IC modeling of both relic regions results in unconstrained temperature and photon index values as well, which can be attributed to the lack of S/N that prohibits the fit to find the best-fit values since 2T model implementation is also unstable and problematic. Assuming the IC and synchrotron emission are due to the same relativistic electron population, we fixed the photon index values of the IC models to the radio spectral indices reported by \citep{jones21}, and we could constrain the temperatures of the 1T+IC model. Both NR and SR region temperatures obtained from the 1T+IC fits agree with the the single temperature fitting (1T) temperature results within 1$\sigma$, and with better constraints. Although all applied models are statistically similarly favourable, the best-fit model describing the X-ray emission from the relic regions is a single temperature thermal plasma with the hint of a non-thermal emission contribution. More data are required to be able investigate the emission characteristics from the relics. The fourth model, IC, was applied to both relic regions to provide a higher limit to the detection signal of a non-thermal emission.

The temperature distribution shown in Figure~\ref{fig:spectralfitNuCH} shows a larger temperature decrease to the north than to the south from the center. While the southern decline may be what is expected for the radial temperature distribution of a non-cool core cluster, abrupt changes in the northern direction may point to a shock front. 

An interesting jump from NR to NR$_{out}$ is also visible in this temperature plot. Although this jump is physically an indication of a cold front, a cold front preceding a shock front at a relic location is not physically motivated. Instead, a non-thermal emission may exist at this location, that drives the temperatures of a single temperature model to assume higher temperatures. The unconstrained temperature values of the \chandra~analysis at the NR$_{out}$ region supports this possibility. Deeper \nustar~data is needed to find the origin of this behaviour, given that it is physical. \xmm~surface brightness analysis finds no evidence of a jump at the northern relic \citep{finner21}. 

Applying Rankine–Hugoniot jump conditions, i.e.; (T$_{post-shock}$/T$_{pre-shock}$ = (5M$^{4}$ + 14M$^{2}$ - 3)/16M$^{2}$) \citep{landau59}, to our best-fit temperatures results from the joint \nustar~and \chandra~spectral fits of the inner and outer region of the relics, we obtain the $\mathcal{M}$ values shown in Table
~\ref{tab:mach} assuming the adiabatic index of an ideal monoatomic gas of $\gamma$=5/3. We cannot give these values for the individual \chandra~fits as the temperature parameters are unconstrained.

\begin{deluxetable}{lccc}
\tabletypesize{\scriptsize}
\tablewidth{0pt} 
\tablecaption{Mach numbers of the candidate shock fronts calculated from the temperature jumps obtained from the spectral fits.
\label{tab:mach}}
\tablehead{\\[-0.95em]
&\multicolumn{2}{c}{Region (kT)}& {$\mathcal{M}$}\\[-0.5em]
&\colhead{{post-shock}}& \colhead{{pre-shock}} &}
\startdata
\\[-0.95em]
{\nustar}& N$_{in}$ (4.37$^{+1.74}_{-1.04}$) & N$_{out}$ (8.51$^{+34.67}_{-5.54}$) & $0.87^{+0.28}_{-1.50}$ \\  
&S$_{in}$ (4.77$^{+0.98}_{-0.67}$) & S$_{out}$ (2.35$^{+2.28}_{-1.34}$)& $1.15^{+0.56}_{-0.91}$ \\
\\[-0.95em]
\hline
\\[-0.95em]
{Joint}& N$_{in}$ (4.07$^{+1.03}_{-0.31}$) & N$_{out}$ (8.82$^{+24.61}_{-5.25}$) & 0.86$^{+0.20}_{-0.86}$ \\ 
&S$_{in}$ (4.76$^{+0.73}_{-0.56}$) & S$_{out}$ (3.38$^{+4.34}_{-2.07}$) & 1.07$^{+0.67}_{-1.34}$ \\
\\[-0.95em]
\enddata
\end{deluxetable}

Our $\mathcal{M}$ results for the southern jump is within 2$\sigma$ of the $\mathcal{M}_{inj}$ reported by \citet{jones21}. However, our northern $\mathcal{M}$ values do not match their results. It is expected given the unphysically high temperatures values we find for the N$_{out}$ region since at these temperatures \nustar~hard band images would have shown enhanced emission at these locations. We note that while \citet{jones21} assume the outer edge of the relics to be where the shock front resides, we assume that location lies close towards the middle of the relics in the radial direction. In addition, GGM images (Figure~\ref{fig:GGM}) are too noisy at the relic locations to make an assessment of such shock location. They also report $\mathcal{M}_{int}$ values that are 1.3 higher than $\mathcal{M}_{intj}$ for the southern jump, which is expected since integrated $\mathcal{M}$ number, $\mathcal{M}_{int}$, is calculated using the emission weighted spectral index distribution where higher Mach numbers have more weight. On the other hand, $\mathcal{M}$ numbers based on the injection spectral index are able to trace the local variations along the relic edge \citep{wittor21}. They also cannot constrain $\mathcal{M}_{int}$ for the northern relic.

A deeper \chandra~exposure is required to draw accurate surface brightness profiles that will provide a stronger motivation for region selections hence a more precise location of the jumps that would help to obtain more accurate $\mathcal{M}$ numbers.

Although there is evidence for a surface brightness jump with a compression factor of C~$\sim$~1.2 for the radial profile drawn from the southern sector in the \xmm~analysis by \citet{finner21}, this jump is found to be unrelated with the relic emission. In addition, short \xmm~exposure does not allow authors to constrain the temperatures at both sides of this jump. Limited \chandra~exposure has also made the detection of shocks difficult \citep{jones21}. These emphasize the requirement for deeper follow up \chandra~and \nustar~observations.

The cross-talk analysis ({\tt nucrossarf}) shows that with the inclusion of PSF contributions from more dominant regions of the cluster, we have a lower IC detection significance from the relic regions than what is found with standard pipeline spectral analysis ({\tt nuproducts}). This is expected as the PSF leakage from the stronger emission regions (such as the Circle region) into the relic regions increase the photon counts of the relic spectra without applying {\tt nucrossarf}, resulting in higher model normalizations hence higher detection significance of any model that would be applied. 

It is also important to note that \nustar~and \chandra~temperature results are in a very good agreement (Figure~\ref{fig:spectralfitNuCH}) where we have moderate S/N; namely NR, NR$_{in}$, Center, SR$_{in}$, and SR regions for the $\sim$2~--~7 keV range.

\section{Conclusion and Future Work}\label{sec:future}
In this work, we analyzed 30~ks raw exposure of \nustar~and 43~ks raw exposure of \chandra~data of a rare double radio relic system \zwcl. We found an indication of a shock front coinciding with the northern radio shock (relic) with both \nustar~and \chandra~spectra, and evidence for weak IC emission at the relic sites using \nustar~data.

Combining the distinct superiorities of these two instruments provides the assessment of a more complete description of the X-ray emission through wider combined bandpass, better capability to constrain multiple temperatures and non-thermal contribution to the signal.

We have established a ground work for the \nustar~follow up observation of this system that is scheduled for 240~ks during Cycle-9. In addition, \chandra~will also observe the system with a deep exposure ($\sim$200~ks) for Cycle-24. These two sets of data together will provide strong claims on the existence and locations of the shock/cold fronts as well as a robust detection or refutation of IC emission at the relic sites. This will be possible not only by increasing the photon statistics of both satellite data on the cluster to constrain the model parameters better, but with higher photon statistics, we will be able to use \nustar's bandpass well beyond 10.0~keV (to at least 15.0~keV) that will enhance the confidence of the possible detection limits of the non-thermal components in the ICM. Furthermore, by obtaining better constraints on the parameters of the 1T+IC models, reliable upper limits on the IC, hence lower limits on the magnetic field strength at relic sites will be driven.

%% IMPORTANT! The old "\acknowledgment" command has be depreciated. It was
%% not robust enough to handle our new dual anonymous review requirements and
%% thus been replaced with the acknowledgment environment. If you try to 
%% compile with \acknowledgment you will get an error print to the screen
%% and in the compiled pdf.
\begin{acknowledgments}

We thank the anonymous referee for contributing to our work with valuable suggestions. AT acknowledges support from NASA NuSTAR Guest Observer grant NNH21ZDA001N-NUSTAR. AT and DRW acknowledges support from NASA ADAP award 80NSSC19K1443. AT thanks Michael McDonald, Duy Hoang, Alex Jones and Ralph P. Kraft for valuable discussions.
This research has made use of data from the \nustar\ mission, a project led by the California Institute of Technology, managed by the Jet Propulsion Laboratory (JPL), and funded by by the National Aeronautics and Space Administration (NASA). In this work, we used the NuSTAR Data Analysis Software (NuSTARDAS) jointly developed by the ASI Science Data Center (ASDC, Italy) and the California Institute of Technology (USA). The data for this research have been obtained from the High Energy Astrophysics Science Archive Research Center (HEASARC), provided by NASA’s Goddard Space Flight Center. 
This paper employs a Chandra dataset, obtained by the Chandra X-ray Observatory, contained in~\dataset[Chandra Data Collection (CDC) 179]{https://doi.org/10.25574/cdc.179}, and software provided by the Chandra X-ray Center (CXC) in the application package CIAO. 
\end{acknowledgments}
\bibliography{zwcl1856}{}

\begin{thebibliography}{}
\expandafter\ifx\csname natexlab\endcsname\relax\def\natexlab#1{#1}\fi
\providecommand{\url}[1]{\href{#1}{#1}}
\providecommand{\dodoi}[1]{doi:~\href{http://doi.org/#1}{\nolinkurl{#1}}}
\providecommand{\doeprint}[1]{\href{http://ascl.net/#1}{\nolinkurl{http://ascl.net/#1}}}
\providecommand{\doarXiv}[1]{\href{https://arxiv.org/abs/#1}{\nolinkurl{https://arxiv.org/abs/#1}}}

\bibitem[{{Biffi} {et~al.}(2016){Biffi}, {Borgani}, {Murante}, {Rasia},
  {Planelles}, {Granato}, {Ragone-Figueroa}, {Beck}, {Gaspari}, \&
  {Dolag}}]{Biffi16}
{Biffi}, V., {Borgani}, S., {Murante}, G., {et~al.} 2016, \apj, 827, 112,
  \dodoi{10.3847/0004-637X/827/2/112}

\bibitem[{{Bonafede} {et~al.}(2012){Bonafede}, {Br{\"u}ggen}, {van Weeren},
  {Vazza}, {Giovannini}, {Ebeling}, {Edge}, {Hoeft}, \& {Klein}}]{bonafede12}
{Bonafede}, A., {Br{\"u}ggen}, M., {van Weeren}, R., {et~al.} 2012, \mnras,
  426, 40, \dodoi{10.1111/j.1365-2966.2012.21570.x}

\bibitem[{{Brunetti} \& {Jones}(2014)}]{brunetti14}
{Brunetti}, G., \& {Jones}, T.~W. 2014, International Journal of Modern Physics
  D, 23, 1430007, \dodoi{10.1142/S0218271814300079}

\bibitem[{{Cash}(1979)}]{cash79}
{Cash}, W. 1979, \apj, 228, 939, \dodoi{10.1086/156922}

\bibitem[{{Dawson}(2013)}]{dawson13}
{Dawson}, W.~A. 2013, \apj, 772, 131, \dodoi{10.1088/0004-637X/772/2/131}

\bibitem[{{de Gasperin} {et~al.}(2014){de Gasperin}, {van Weeren},
  {Br{\"u}ggen}, {Vazza}, {Bonafede}, \& {Intema}}]{degasperin14}
{de Gasperin}, F., {van Weeren}, R.~J., {Br{\"u}ggen}, M., {et~al.} 2014,
  \mnras, 444, 3130, \dodoi{10.1093/mnras/stu1658}

\bibitem[{{Finner} {et~al.}(2021){Finner}, {HyeongHan}, {Jee}, {Wittman},
  {Forman}, {van Weeren}, {Golovich}, {Dawson}, {Jones}, {de Gasperin}, \&
  {Jones}}]{finner21}
{Finner}, K., {HyeongHan}, K., {Jee}, M.~J., {et~al.} 2021, \apj, 918, 72,
  \dodoi{10.3847/1538-4357/ac0d00}

\bibitem[{{Gabici} \& {Blasi}(2003)}]{gabici03}
{Gabici}, S., \& {Blasi}, P. 2003, \apj, 583, 695, \dodoi{10.1086/345429}

\bibitem[{{Harrison} {et~al.}(2013){Harrison}, {Craig}, {Christensen},
  {Hailey}, {Zhang}, {Boggs}, {Stern}, {Cook}, {Forster}, {Giommi},
  {Grefenstette}, {Kim}, {Kitaguchi}, {Koglin}, {Madsen}, {Mao}, {Miyasaka},
  {Mori}, {Perri}, {Pivovaroff}, {Puccetti}, {Rana}, {Westergaard}, {Willis},
  {Zoglauer}, {An}, {Bachetti}, {Barri{\`e}re}, {Bellm}, {Bhalerao},
  {Brejnholt}, {Fuerst}, {Liebe}, {Markwardt}, {Nynka}, {Vogel}, {Walton},
  {Wik}, {Alexander}, {Cominsky}, {Hornschemeier}, {Hornstrup}, {Kaspi},
  {Madejski}, {Matt}, {Molendi}, {Smith}, {Tomsick}, {Ajello}, {Ballantyne},
  {Balokovi{\'c}}, {Barret}, {Bauer}, {Blandford}, {Brandt}, {Brenneman},
  {Chiang}, {Chakrabarty}, {Chenevez}, {Comastri}, {Dufour}, {Elvis}, {Fabian},
  {Farrah}, {Fryer}, {Gotthelf}, {Grindlay}, {Helfand}, {Krivonos}, {Meier},
  {Miller}, {Natalucci}, {Ogle}, {Ofek}, {Ptak}, {Reynolds}, {Rigby},
  {Tagliaferri}, {Thorsett}, {Treister}, \& {Urry}}]{harrison13}
{Harrison}, F.~A., {Craig}, W.~W., {Christensen}, F.~E., {et~al.} 2013, \apj,
  770, 103, \dodoi{10.1088/0004-637X/770/2/103}

\bibitem[{{Jones} {et~al.}(2021){Jones}, {de Gasperin}, {Cuciti}, {Hoang},
  {Botteon}, {Br{\"u}ggen}, {Brunetti}, {Finner}, {Forman}, {Jones}, {Kraft},
  {Shimwell}, \& {van Weeren}}]{jones21}
{Jones}, A., {de Gasperin}, F., {Cuciti}, V., {et~al.} 2021, \mnras, 505, 4762,
  \dodoi{10.1093/mnras/stab1443}

\bibitem[{{Kalberla} {et~al.}(2005){Kalberla}, {Burton}, {Hartmann}, {Arnal},
  {Bajaja}, {Morras}, \& {P{\"o}ppel}}]{kalberla05}
{Kalberla}, P.~M.~W., {Burton}, W.~B., {Hartmann}, D., {et~al.} 2005, \aap,
  440, 775, \dodoi{10.1051/0004-6361:20041864}

\bibitem[{{Landau} \& {Lifshitz}(1959)}]{landau59}
{Landau}, L.~D., \& {Lifshitz}, E.~M. 1959, {Fluid mechanics}

\bibitem[{{Markevitch} \& {Vikhlinin}(2007)}]{markevitch07}
{Markevitch}, M., \& {Vikhlinin}, A. 2007, \physrep, 443, 1,
  \dodoi{10.1016/j.physrep.2007.01.001}

\bibitem[{{Planck Collaboration VIII} {et~al.}(2011){Planck Collaboration
  VIII}, {Ade}, {Aghanim}, {Arnaud}, {Ashdown}, {Aumont}, {Baccigalupi},
  {Balbi}, {Banday}, {Barreiro}, {Bartelmann}, {Bartlett}, {Battaner},
  {Battye}, {Benabed}, {Beno{\^\i}t}, {Bernard}, {Bersanelli}, {Bhatia},
  {Bock}, {Bonaldi}, {Bond}, {Borrill}, {Bouchet}, {Brown}, {Bucher},
  {Burigana}, {Cabella}, {Cantalupo}, {Cardoso}, {Carvalho}, {Catalano},
  {Cay{\'o}n}, {Challinor}, {Chamballu}, {Chary}, {Chiang}, {Chiang}, {Chon},
  {Christensen}, {Churazov}, {Clements}, {Colafrancesco}, {Colombi}, {Couchot},
  {Coulais}, {Crill}, {Cuttaia}, {da Silva}, {Dahle}, {Danese}, {Davis}, {de
  Bernardis}, {de Gasperis}, {de Rosa}, {de Zotti}, {Delabrouille}, {Delouis},
  {D{\'e}sert}, {Dickinson}, {Diego}, {Dolag}, {Dole}, {Donzelli}, {Dor{\'e}},
  {D{\"o}rl}, {Douspis}, {Dupac}, {Efstathiou}, {Eisenhardt}, {En{\ss}lin},
  {Feroz}, {Finelli}, {Flores-Cacho}, {Forni}, {Fosalba}, {Frailis},
  {Franceschi}, {Fromenteau}, {Galeotta}, {Ganga}, {G{\'e}nova-Santos},
  {Giard}, {Giardino}, {Giraud-H{\'e}raud}, {Gonz{\'a}lez-Nuevo},
  {Gonz{\'a}lez-Riestra}, {G{\'o}rski}, {Grainge}, {Gratton}, {Gregorio},
  {Gruppuso}, {Harrison}, {Hein{\"a}m{\"a}ki}, {Henrot-Versill{\'e}},
  {Hern{\'a}ndez-Monteagudo}, {Herranz}, {Hildebrandt}, {Hivon}, {Hobson},
  {Holmes}, {Hovest}, {Hoyland}, {Huffenberger}, {Hurier}, {Hurley-Walker},
  {Jaffe}, {Jones}, {Juvela}, {Keih{\"a}nen}, {Keskitalo}, {Kisner}, {Kneissl},
  {Knox}, {Kurki-Suonio}, {Lagache}, {Lamarre}, {Lasenby}, {Laureijs},
  {Lawrence}, {Le Jeune}, {Leach}, {Leonardi}, {Li}, {Liddle}, {Lilje},
  {Linden-V{\o}rnle}, {L{\'o}pez-Caniego}, {Lubin}, {Mac{\'\i}as-P{\'e}rez},
  {MacTavish}, {Maffei}, {Maino}, {Mandolesi}, {Mann}, {Maris}, {Marleau},
  {Mart{\'\i}nez-Gonz{\'a}lez}, {Masi}, {Matarrese}, {Matthai}, {Mazzotta},
  {Mei}, {Meinhold}, {Melchiorri}, {Melin}, {Mendes}, {Mennella}, {Mitra},
  {Miville-Desch{\^e}nes}, {Moneti}, {Montier}, {Morgante}, {Mortlock},
  {Munshi}, {Murphy}, {Naselsky}, {Nati}, {Natoli}, {Netterfield},
  {N{\o}rgaard-Nielsen}, {Noviello}, {Novikov}, {Novikov}, {Olamaie},
  {Osborne}, {Pajot}, {Pasian}, {Patanchon}, {Pearson}, {Perdereau}, {Perotto},
  {Perrotta}, {Piacentini}, {Piat}, {Pierpaoli}, {Piffaretti}, {Plaszczynski},
  {Pointecouteau}, {Polenta}, {Ponthieu}, {Poutanen}, {Pratt}, {Pr{\'e}zeau},
  {Prunet}, {Puget}, {Rachen}, {Reach}, {Rebolo}, {Reinecke}, {Renault},
  {Ricciardi}, {Riller}, {Ristorcelli}, {Rocha}, {Rosset},
  {Rubi{\~n}o-Mart{\'\i}n}, {Rusholme}, {Saar}, {Sandri}, {Santos}, {Saunders},
  {Savini}, {Schaefer}, {Scott}, {Seiffert}, {Shellard}, {Smoot}, {Stanford},
  {Starck}, {Stivoli}, {Stolyarov}, {Stompor}, {Sudiwala}, {Sunyaev}, {Sutton},
  {Sygnet}, {Taburet}, {Tauber}, {Terenzi}, {Toffolatti}, {Tomasi}, {Torre},
  {Tristram}, {Tuovinen}, {Valenziano}, {Vibert}, {Vielva}, {Villa},
  {Vittorio}, {Wade}, {Wandelt}, {Weller}, {White}, {White}, {Yvon}, {Zacchei},
  \& {Zonca}}]{planck11}
{Planck Collaboration VIII}, {Ade}, P.~A.~R., {Aghanim}, N., {et~al.} 2011,
  \aap, 536, A8, \dodoi{10.1051/0004-6361/201116459}

\bibitem[{{Planck Collaboration XXVII} {et~al.}(2016){Planck Collaboration
  XXVII}, {Ade}, {Aghanim}, {Arnaud}, {Ashdown}, {Aumont}, {Baccigalupi},
  {Banday}, {Barreiro}, {Barrena}, {Bartlett}, {Bartolo}, {Battaner}, {Battye},
  {Benabed}, {Beno{\^\i}t}, {Benoit-L{\'e}vy}, {Bernard}, {Bersanelli},
  {Bielewicz}, {Bikmaev}, {B{\"o}hringer}, {Bonaldi}, {Bonavera}, {Bond},
  {Borrill}, {Bouchet}, {Bucher}, {Burenin}, {Burigana}, {Butler}, {Calabrese},
  {Cardoso}, {Carvalho}, {Catalano}, {Challinor}, {Chamballu}, {Chary},
  {Chiang}, {Chon}, {Christensen}, {Clements}, {Colombi}, {Colombo}, {Combet},
  {Comis}, {Couchot}, {Coulais}, {Crill}, {Curto}, {Cuttaia}, {Dahle},
  {Danese}, {Davies}, {Davis}, {de Bernardis}, {de Rosa}, {de Zotti},
  {Delabrouille}, {D{\'e}sert}, {Dickinson}, {Diego}, {Dolag}, {Dole},
  {Donzelli}, {Dor{\'e}}, {Douspis}, {Ducout}, {Dupac}, {Efstathiou},
  {Eisenhardt}, {Elsner}, {En{\ss}lin}, {Eriksen}, {Falgarone}, {Fergusson},
  {Feroz}, {Ferragamo}, {Finelli}, {Forni}, {Frailis}, {Fraisse}, {Franceschi},
  {Frejsel}, {Galeotta}, {Galli}, {Ganga}, {G{\'e}nova-Santos}, {Giard},
  {Giraud-H{\'e}raud}, {Gjerl{\o}w}, {Gonz{\'a}lez-Nuevo}, {G{\'o}rski},
  {Grainge}, {Gratton}, {Gregorio}, {Gruppuso}, {Gudmundsson}, {Hansen},
  {Hanson}, {Harrison}, {Hempel}, {Henrot-Versill{\'e}},
  {Hern{\'a}ndez-Monteagudo}, {Herranz}, {Hildebrandt}, {Hivon}, {Hobson},
  {Holmes}, {Hornstrup}, {Hovest}, {Huffenberger}, {Hurier}, {Jaffe}, {Jaffe},
  {Jin}, {Jones}, {Juvela}, {Keih{\"a}nen}, {Keskitalo}, {Khamitov}, {Kisner},
  {Kneissl}, {Knoche}, {Kunz}, {Kurki-Suonio}, {Lagache}, {Lamarre}, {Lasenby},
  {Lattanzi}, {Lawrence}, {Leonardi}, {Lesgourgues}, {Levrier}, {Liguori},
  {Lilje}, {Linden-V{\o}rnle}, {L{\'o}pez-Caniego}, {Lubin},
  {Mac{\'\i}as-P{\'e}rez}, {Maggio}, {Maino}, {Mak}, {Mandolesi}, {Mangilli},
  {Martin}, {Mart{\'\i}nez-Gonz{\'a}lez}, {Masi}, {Matarrese}, {Mazzotta},
  {McGehee}, {Mei}, {Melchiorri}, {Melin}, {Mendes}, {Mennella}, {Migliaccio},
  {Mitra}, {Miville-Desch{\^e}nes}, {Moneti}, {Montier}, {Morgante},
  {Mortlock}, {Moss}, {Munshi}, {Murphy}, {Naselsky}, {Nastasi}, {Nati},
  {Natoli}, {Netterfield}, {N{\o}rgaard-Nielsen}, {Noviello}, {Novikov},
  {Novikov}, {Olamaie}, {Oxborrow}, {Paci}, {Pagano}, {Pajot}, {Paoletti},
  {Pasian}, {Patanchon}, {Pearson}, {Perdereau}, {Perotto}, {Perrott},
  {Perrotta}, {Pettorino}, {Piacentini}, {Piat}, {Pierpaoli}, {Pietrobon},
  {Plaszczynski}, {Pointecouteau}, {Polenta}, {Pratt}, {Pr{\'e}zeau}, {Prunet},
  {Puget}, {Rachen}, {Reach}, {Rebolo}, {Reinecke}, {Remazeilles}, {Renault},
  {Renzi}, {Ristorcelli}, {Rocha}, {Rosset}, {Rossetti}, {Roudier}, {Rozo},
  {Rubi{\~n}o-Mart{\'\i}n}, {Rumsey}, {Rusholme}, {Rykoff}, {Sandri}, {Santos},
  {Saunders}, {Savelainen}, {Savini}, {Schammel}, {Scott}, {Seiffert},
  {Shellard}, {Shimwell}, {Spencer}, {Stanford}, {Stern}, {Stolyarov},
  {Stompor}, {Streblyanska}, {Sudiwala}, {Sunyaev}, {Sutton}, {Suur-Uski},
  {Sygnet}, {Tauber}, {Terenzi}, {Toffolatti}, {Tomasi}, {Tramonte},
  {Tristram}, {Tucci}, {Tuovinen}, {Umana}, {Valenziano}, {Valiviita}, {Van
  Tent}, {Vielva}, {Villa}, {Wade}, {Wandelt}, {Wehus}, {White}, {Wright},
  {Yvon}, {Zacchei}, \& {Zonca}}]{planck16}
{Planck Collaboration XXVII}, {Ade}, P.~A.~R., {Aghanim}, N., {et~al.} 2016,
  \aap, 594, A27, \dodoi{10.1051/0004-6361/201525823}

\bibitem[{{Pratt} {et~al.}(2009){Pratt}, {Croston}, {Arnaud}, \&
  {B{\"o}hringer}}]{pratt09}
{Pratt}, G.~W., {Croston}, J.~H., {Arnaud}, M., \& {B{\"o}hringer}, H. 2009,
  \aap, 498, 361, \dodoi{10.1051/0004-6361/200810994}

\bibitem[{Ryu {et~al.}(2003)Ryu, Kang, Hallman, \& Jones}]{ryu03}
Ryu, D., Kang, H., Hallman, E., \& Jones, T.~W. 2003, The Astrophysical
  Journal, 593, 599–610, \dodoi{10.1086/376723}

\bibitem[{{Sanders} {et~al.}(2016){Sanders}, {Fabian}, {Russell}, {Walker}, \&
  {Blundell}}]{sanders16}
{Sanders}, J.~S., {Fabian}, A.~C., {Russell}, H.~R., {Walker}, S.~A., \&
  {Blundell}, K.~M. 2016, \mnras, 460, 1898, \dodoi{10.1093/mnras/stw1119}

\bibitem[{{Sarazin} {et~al.}(2016){Sarazin}, {Finoguenov}, {Wik}, \&
  {Clarke}}]{Sarazin16}
{Sarazin}, C.~L., {Finoguenov}, A., {Wik}, D.~R., \& {Clarke}, T.~E. 2016,
  arXiv e-prints, arXiv:1606.07433.
\newblock \doarXiv{1606.07433}

\bibitem[{{T{\"u}mer} {et~al.}(2022){T{\"u}mer}, {Wik}, {Gaspari}, {Akamatsu},
  {Westergaard}, {Tombesi}, \& {Ercan}}]{tumer22}
{T{\"u}mer}, A., {Wik}, D.~R., {Gaspari}, M., {et~al.} 2022, \apj, 930, 83,
  \dodoi{10.3847/1538-4357/ac61de}

\bibitem[{{T{\"u}mer} {et~al.}(2023){T{\"u}mer}, {Wik}, {Zhang}, {Hoang},
  {Gaspari}, {van Weeren}, {Rudnick}, {Stuardi}, {Mernier}, {Simionescu},
  {Rojas Bolivar}, {Kraft}, {Akamatsu}, \& {de Plaa}}]{tumer23}
{T{\"u}mer}, A., {Wik}, D.~R., {Zhang}, X., {et~al.} 2023, \apj, 942, 79,
  \dodoi{10.3847/1538-4357/aca1b5}

\bibitem[{{van Weeren} {et~al.}(2019){van Weeren}, {de Gasperin}, {Akamatsu},
  {Br{\"u}ggen}, {Feretti}, {Kang}, {Stroe}, \& {Zandanel}}]{weeren19}
{van Weeren}, R.~J., {de Gasperin}, F., {Akamatsu}, H., {et~al.} 2019, \ssr,
  215, 16, \dodoi{10.1007/s11214-019-0584-z}

\bibitem[{Vazza {et~al.}(2012)Vazza, Brüggen, van Weeren, Bonafede, Dolag, \&
  Brunetti}]{vazza12}
Vazza, F., Brüggen, M., van Weeren, R., {et~al.} 2012, Monthly Notices of the
  Royal Astronomical Society, 421, 1868,
  \dodoi{10.1111/j.1365-2966.2011.20160.x}

\bibitem[{{Vikhlinin} {et~al.}(2009){Vikhlinin}, {Burenin}, {Ebeling},
  {Forman}, {Hornstrup}, {Jones}, {Kravtsov}, {Murray}, {Nagai}, {Quintana}, \&
  {Voevodkin}}]{vikhlinin09}
{Vikhlinin}, A., {Burenin}, R.~A., {Ebeling}, H., {et~al.} 2009, \apj, 692,
  1033, \dodoi{10.1088/0004-637X/692/2/1033}

\bibitem[{{Wik} {et~al.}(2014){Wik}, {Hornstrup}, {Molendi}, {Madejski},
  {Harrison}, {Zoglauer}, {Grefenstette}, {Gastaldello}, {Madsen},
  {Westergaard}, {Ferreira}, {Kitaguchi}, {Pedersen}, {Boggs}, {Christensen},
  {Craig}, {Hailey}, {Stern}, \& {Zhang}}]{wik14}
{Wik}, D.~R., {Hornstrup}, A., {Molendi}, S., {et~al.} 2014, \apj, 792, 48,
  \dodoi{10.1088/0004-637X/792/1/48}

\bibitem[{{Wilms} {et~al.}(2000){Wilms}, {Allen}, \& {McCray}}]{wilms00}
{Wilms}, J., {Allen}, A., \& {McCray}, R. 2000, \apj, 542, 914,
  \dodoi{10.1086/317016}

\bibitem[{{Wittor} {et~al.}(2021){Wittor}, {Ettori}, {Vazza}, {Rajpurohit},
  {Hoeft}, \& {Dom{\'\i}nguez-Fern{\'a}ndez}}]{wittor21}
{Wittor}, D., {Ettori}, S., {Vazza}, F., {et~al.} 2021, \mnras, 506, 396,
  \dodoi{10.1093/mnras/stab1735}

\end{thebibliography}
\bibliographystyle{aasjournal}

\end{document}